\documentclass[11pt,a4paper]{article}

\usepackage{jheppub}

\usepackage{amssymb}
\usepackage{amsfonts}
\usepackage{graphicx}
\usepackage{epstopdf}
\usepackage{dcolumn}
\usepackage{amsmath}
\usepackage{latexsym,bm}
\usepackage{amsthm}
\usepackage{slashed}
\usepackage{float}
\usepackage{color}
\usepackage{url}
\usepackage{longtable}
\usepackage{subfig}
\usepackage{tikz}
\usepackage[all,cmtip]{xy}
\usepackage{hyperref}

\usepackage{subfig}

\newcommand \xoverline[2][0.75]{
    \sbox{\myboxA}{$\m@th#2$}
    \setbox\myboxB\null
    \ht\myboxB=\ht\myboxA
    \dp\myboxB=\dp\myboxA
    \wd\myboxB=#1\wd\myboxA
    \sbox\myboxB{$\m@th\overline{\copy\myboxB}$}
    \setlength\mylenA{\the\wd\myboxA}
    \addtolength\mylenA{-\the\wd\myboxB}
    \ifdim\wd\myboxB<\wd\myboxA
       \rlap{\hskip 0.5\mylenA\usebox\myboxB}{\usebox\myboxA}%
    \else
        \hskip -0.5\mylenA\rlap{\usebox\myboxA}{\hskip 0.5\mylenA\usebox\myboxB}%
    \fi}
\makeatother

\newcommand{\ba}{\begin{aligned}}
\newcommand{\ea}{\end{aligned}}
\def \be {\begin{equation}}
\def \ee {\end{equation}}
\def \bsp {\begin{split}}
\def \esp {\end{split}}
\def \bea {\begin{eqnarray}}
\def \eea {\end{eqnarray}}

\def \bp{\begin{pmatrix}}
\def\ep{\end{pmatrix}}

\def\R{\mathbb{R}}
\def\N{\mathcal{N}}
\def\P{\mathbb{P}}
\def\C{\mathbb{C}}
\def\Z{\mathbb{Z}}

\title {Gauging Discrete Symmetries of $T_N$-theories in Five Dimensions}

\preprint{\today \hspace*{0.1in} }

\usepackage{amsmath}
\DeclareMathOperator{\Tr}{Tr}
\usepackage{physics}

\usepackage{float}

\usepackage{graphicx}
\graphicspath{ {./images/} }
\usepackage{adjustbox}

\usepackage{etoolbox} 
    \makeatletter
    \patchcmd{\maketitle}{\@fpheader}{}{}{}
    \makeatother
    
\preprint{KCL-PH-TH/2021-78}

\author[a,b]{Bobby Acharya,} \author[c]
{Neil Lambert}\author[a,b]{Marwan Najjar,}\author[d]{Eirik Eik Svanes}\author[a]{and Jiahua Tian}
                                
 \affiliation[a]{ICTP\\
 Strada Costiera 11 34151, Trieste, Italy}
  \affiliation[b]{ Department of Physics\\ King's College London\\
 The Strand WC2R 2LS, London, UK}
  \affiliation[c]{ Department of Mathematics\\ King's College London\\
 The Strand WC2R 2LS, London, UK}
  \affiliation[d]{ Department of Mathematics and Physics\\ University of Stavanger\\ Kristine Bonnevies vei 22, 4021
Stavanger, Norway}
 
 \emailAdd{bacharya@ictp.it} \emailAdd{neil.lambert@kcl.ac.uk} \emailAdd{mnajjar1@ictp.it} \emailAdd{eirik.e.svanes@uis.no}
 \emailAdd{jtian@ictp.it}

\abstract{We study the the gauging of a discrete $\Z_3$ symmetry in the five-dimensional superconformal $T_N$ theories. 
We argue that this leads to an infinite sequence of five-dimensional superconformal theories with either $E_6 \times SU(N)$ or $SU(3)\times SU(N)$ global symmetry group. In the $M$-theory realisation of $T_N$ theories as residing at the origin in the Calabi-Yau orbifolds ${\C^3 \over {\Z_N \times \Z_N}}$ we identify the $\Z_3$ symmetry geometrically and the new theories arise from $M$-theory on the the non-Abelian orbifolds  $({\C^3 \over {\Z_N \times \Z_N}})/{\Z_3}$. On the other hand, in the $(p,q)$ 5-brane web description in Type IIB theory, the symmetry combines the $U$-duality symmetry with a rotation in space, defining a so-called $U$-fold background, where the $E_6$ symmetry is manifest.}


\begin{document}

\maketitle

\section{Introduction}

Gauge theories are the building blocks of modern theoretical physics. They have well documented applications to condensed physics, particle physics and, {\it via} the AdS/CFT correspondence, gravity. Traditionally gauge theories are constructed as Lagrangian theories based on the Yang-Mills action (or possibly the Chern-Simons action in three-dimensions). As such perturbative renormalisability, vacuum stability and UV completeness restricts us to gauge theories in four or less dimensions. 

Superstring/$M$-theory has taught us that there are much broader classes of gauge theories, ones which are not necessarily based on Lagrangians. Furthermore such theories can exist in five or six dimensions. These theories typically arise in Superstring/$M$-theory as degrees of freedom localised on either $M$5-branes or at {\it very particular} kinds of singularities, where one can decouple bulk gravity. As a result the gravitational degrees of freedom decouple. Under the correct conditions, these localised dynamical degrees of freedom are controlled by a conformal field theory which could also have a gauge theory interpretation. In this paper we will study particular kinds of singularities in $M$-theory which give rise to five-dimensional superconformal theories.
 
 A wide class of five-dimensional theories are believed to arise in M-theory spacetimes of the form ${\mathbb R}^5\times X$ where $X$ is a non-compact, Calabi-Yau 3-fold with certain kinds of singularities. This idea goes back to the original work of \cite{Morrison:1996xf, Seiberg:1996bd, Intriligator:1997pq}. There it was argued that if the singularities of $X$ arise from the collapse of a four-dimensional submanifold to a point, then the degrees of freedom of a superconformal field theory are localised there. Subsequently there has been a huge amount of literature on this correspondence \cite{Xie_2017, Jefferson:2017ahm, Jefferson_2018, Closset:2018bjz, Closset:2020afy, Closset:2019juk, Closset_2021, Bhardwaj:2018vuu, Bhardwaj:2018yhy, Bhardwaj:2019jtr, Bhardwaj:2019ngx, Bhardwaj:2019xeg, Bhardwaj:2020gyu, Bhardwaj:2020ruf, Bhardwaj:2020avz, Apruzzi_2019, Apruzzi:2019enx, Apruzzi:2019kgb, Apruzzi:2019vpe, Eckhard:2020jyr, Saxena:2020ltf}.

Although M-theory is not well understood, enough is known that allows one to gain some understanding about the five-dimensional theory which governs the low energy dynamics associated to the singularity. In particular there are parallel algebra-geometric and gauge theory desingularisations of the singularities that can be used to analyse the system. From the geometrical point of view many such singular Calabi-Yau manifolds admit smooth resolutions. As such the low energy description becomes that of an abelian gauge theory obtained by considering the eleven-dimensional supergravity approximation along with wrapped branes, on the resolved spacetime.  On the other hand, from the gauge theory point of view, this corresponds to looking at the dynamics on the Coulomb branch, where the gauge group is broken to its Cartan subalgebra. There are also Higgs' and mixed Coulomb/Higgs'  branches whose understanding is more involved and will not be addressed here. 

\subsection{Metrics on Singularities}

Most of the literature on the singular Calabi-Yau threefolds which arise focuses on the algebraic geometry perspective of the singularity. However, $M$-theory is described by a field theory at low energies and the spacetime metric is one of its fields. One can therefore ask about the Calabi-Yau metric of the singular spacetime. This brief subsection describes issues that arise in considering the metric and defines the class of metrics that we will henceforth consider. Since, after decoupling gravity, one is really dealing with a non-compact space, $X$ one has to consider the asymptotic form of the metric at ``infinity". If infinity is approached as a single coordinate of $X$ grows, then the metric can be brought into a form in which the leading terms are
\be
ds^2 \sim dr^2 + r^{\alpha} ds_5^2 (\Sigma)\ ,
\ee
where $ds_5^2$ is the metric on the boundary 5-manifold $\Sigma$ and $\alpha$ is a positive scaling dimension. Notice that, only for $\alpha=2$ does the Ricci tensor, which is zero for a Calabi-Yau metric, remain invariant under a constant rescaling of the $r$ coordinate. Therefore, for $\alpha=2$, the equations of motion have this scaling symmetry which implies that, semi-classically at least, there is a global scaling symmetry of the decoupled system. The case $\alpha=2$ is known as a conically singular metric. A natural, minimal assumption is that a five-dimensional superconformal field theory arises from conically singular Calabi-Yau {\it metrics}. Note that conically singular Ricci flat metrics have been known to be associated with interesting and often conformal field theories in superstring/$M$-theory in a variety of other contexts \cite{Morrison:1998cs, Acharya:1998db, Atiyah:2001qf, Acharya:2001gy}. Also note that this assumption does not preclude SCFT's arising from other types of metric singularities. 

Another aspect of metric singularities concerns the limit one has in mind when decoupling gravity (and other degrees of freedom) completely from the system. In general relativity and presumably also $M$ theory, when one specifies a spacetime background, one usually requires that the spacetime is geodesically complete, at least outside horizons or cosmological singularities. If not, then one has the problem of what happens in the incomplete regions of the space.
For the kinds of singularities that arise in our applications of $M$ theory, one typically resolves the singularities of $X$ to produce a smooth manifold, hence one would require a {\it complete Ricci flat Calabi-Yau metric} on the whole of this space. One is therefore led to postulate {\it complete, asymptotically conical (AC) Calabi-Yau metrics}.
If one considers incomplete metrics, {\it e.g.} defined in the neighbourhood of a submanifold for instance, then it is somewhat unclear how to decouple degrees of freedom residing on the boundary of the region, because the metric has to be modified to a non-Ricci flat geometry there. Henceforth, we will only consider this class of AC Calabi-Yau metrics. We note that it has been claimed that one obtains superconformal field theories whenever $X$ is a Calabi-Yau threefold with  an isolated canonical singularity \cite{Xie_2017}, though it is also known that {\it not} all such spaces can admit conical Calabi-Yau metrics \cite{Gauntlett:2006vf, Collins:2015qsb}.
More generally  it would be interesting to understand the  necessary and sufficient conditions for an isolated singularity to lead to a non-trivial superconformal field theory. 

A large class of AC Calabi-Yau metrics can be generated by considering crepant resolutions of locally flat orbifolds of the form ${\mathbb{C}^3}/\Gamma$, where $\Gamma$ is any finite subgroup of $SU(3)$. One establishes the existence of AC Calabi-Yau metrics on a given crepant resolution $X_{\Gamma} = \widetilde{\mathbb{C}^3/\Gamma}$ as follows. Roan, following work of Ito, Markushevich and others, proved that at least one crepant resolution exists for every $\Gamma$ in $SU(3)$ \cite{doi:10.1142/S0129167X95000043, Markushevich1997, Roan1996MinimalRO}. Then, some twenty years later, van Coevering \cite{Vancoevering} proved that ${\it given}$ a conical Calabi-Yau metric on a space $X_0$  which admits a smooth crepant resolution $X$, there exists a smooth, complete AC Calabi-Yau metric on $X$. Other examples of metrically conical singularities are known to arise in the context of the AdS/CFT correspondence \cite{Acharya:1998db, Gauntlett:2004yd} and these examples are all toric Calabi-Yau's. Asides from these, very few other examples are known. The examples we will study are not toric, though, as we will see they are related to toric examples by a sort of gauging.

A particularly interesting class of five-dimensional SCFT's are the so called $T_N$ theories. In $M$-theory these arise from orbifolds in which $\Gamma$ is a particular ${\mathbb{Z}_N}\times{\mathbb{Z}_N}$ subgroup of $SU(3)$. These conformal theories exhibit a variety of interesting phenomena, such as $SU(N)^3$ exact global symmetries and large rank gauge symmetries along their Coulomb branch for large $N$. For $N=3$ the global symmetry is actually enhanced to $E_6$. We will show that these theories admit a ${\mathbb{Z}_3}$ symmetry, permuting the three $SU(N)$ global symmetries and, that, gauging the symmetry {\it requires} the introduction of additional degrees of freedom. For $N$ congruent to $0$ mod $3$ we will show that there is a global symmetry group which is $E_6 \times SU(N)$. We compute the rank of the gauge group along the Coulomb branch and also use the $(p, q)$ 5-brane web picture of $T_N$ theories to show that in Type IIB theory, gauging the $\Z_3$ symmetry is equivalent to gauging the order three subgroup of the $SL(2,Z)$ strong-weak coupling duality symmetry combined with a rotation in space {\it i.e.} a so-called $S$-fold. This leads to the introduction of additional non-local 7-branes fixed at the origin and hence to exotic flavour symmetries. The ${\mathbb Z}_3$ gauging only exists for a fixed value of the axio-dilaton, so there is no weak coupling limit, and as such one doesn't expect a Lagrangian description.

The rest of this paper is organised as follows. In section \ref{sec:gauge theories} we provide a review of basic features of five-dimensional gauge theories and their construction from M-theory, including the so-called $T_N$ theories. In Section \ref{sec:Newfive-dimensional} we describe the singular Calabi-Yau orbifolds, particular resolutions and how the gauge theory data may be understood group theoretically. In Section \ref{sec:HS} we give a description of ${\mathbb C}^3/\Delta(3N^2)$ and its resolution as a hypersurface embedding which gives further insight into the nature of the resulting gauge theory.  In section \ref{sec:brane_web} we present a dual 5-brane web construction of the five-dimension field theories in terms of a S-fold of familiar 5-brane webs.  In Section \ref{sec:curve} we construct the five-dimensional Seiberg-Witten curve. Finally we give our conclusions in Section \ref{sec:conclusions}.

\section{Five-Dimensional Gauge Theories From M-Theory}\label{sec:gauge theories}

In this section we wish to review various aspects of five-dimensional gauge theories and their construction from M-theory.

\subsection{Low Energy Lagrangian Descriptions}

Minimally supersymmetric gauge theories in five dimensional Minkowski spacetime have
eight real supercharges which transform as a doublet under the R-symmetry group
$SU(2)_R$. The global symmetry of the theory therefore contains the factors
\begin{equation}
SO(1,4)\times SU(2)_R.
\end{equation}
For a compact and connected gauge group $G$ with Lie algebra $\mathfrak{g}$ and rank
$r$, there are two types of supermultiplets which are often considered. Vector
multiplets transform in the adjoint of $G$ and consist of a gauge field
$\{A_{\mu}^{i}; \ i = 1, ..., dim(\mathfrak{g})\}$, a  real scalar $\{\phi^{i}\}$,
and symplectic Majorana spinors $\{\lambda^{i}\}$, {\it {\it i.e.}}
$\{(A_{\mu}^{i},\phi^{i},\lambda^{i})\}$. They transform under the global symmetry
above as $(\textbf{5},\textbf{1}), (\textbf{1},\textbf{1}),
(\textbf{4},\textbf{2})$, respectively. Matter multiplets that transform in an
arbitrary representation $\textbf{R}$ of $G$ are hypermultiplets (HM's). HM's
consist of an $SU(2)_R$ doublet of complex scalars $\{h^{\alpha}; \alpha = 1, ...,
dim(\textbf{R})\}$ and  spinors $\{\psi^{\alpha}\}$ each belongs to
$(\textbf{4},\textbf{2})$. On top of the global symmetries above, the space of HM's
can also be acted upon by various flavour symmetry groups,   denoted as
$G_{F}^{IR}$. 

Because self-dual Yang-Mills instantons on ${\mathbb{R}^4}$ are solitonic particles in
${\mathbb{R}^{1,4}}$, there exists a conserved topological current \cite{Jefferson_2018}, whose
charge is the instanton number. Associated to this is a $U(1)_T$ symmetry group.
All of the fields we mentioned above are uncharged with respect to $U(1)_T$ but
non-perturbative sectors of theory can carry non-trivial charges. Thus, the global
symmetries are at least
\begin{equation}
    SO(1,4) \times SU(2)_R \times G_{F}^{IR}\times U(1)_T.
\end{equation}

The gauge coupling $g$ is an irrelevant parameter, meaning that the theory is
infrared (IR) free. Thus, in the IR, one may study the Coulomb branch (CB) along
which the adjoint scalar $\{\phi^{a}\}$ has a non-zero expectation value and Higgses
$G$ to its maximal torus. The classical CB moduli space is ${\mathbb{R}^r}/W(G)$ where
$W(G)$ is the Weyl group of $G$. At a generic point, we have a $(U(1))^{r}$ Abelian
gauge theory.The low-energy effective action  , like the case of
Seiberg-Witten theory, is determined through a pre-potential
$\mathcal{F}=\mathcal{F}(\phi^{a})$. The low-energy effective action without any matter fields is given by
\begin{equation}
    \mathcal{L}_{eff} = G_{ab}\  d\phi^{a}\wedge \ast d\phi^{b} + G_{ab} \
F^{a}\wedge \ast F^{b} + \frac{c_{abc}}{24\pi^{2}} \ A^{a}\wedge F^{b}\wedge
F^{c}.
\end{equation}
The metric $G_{ab}$ and the constant $c_{abc}$ are determined through the
prepotential by
\begin{equation}
    G_{ab} = \partial_{a}\partial_{b}\mathcal{F}, \ \ c_{abc} =
\partial_{a}\partial_{b}\partial_{c}\mathcal{F}.
\end{equation}
The generic form of the prepotential of a simple gauge group $G$, and when matter
fields are added can be written as\footnote{In fact, we are talking about the
extended coulomb branch (ECB) as we include the masses $m_{f}$.}
\begin{equation}\label{mathcalF}
\begin{split}
    \mathcal{F}(\phi^{a}, m_{f}) &= \frac{1}{2g^{2}} c_{ab} \phi^{a}\phi^{b} +
\frac{\kappa}{6}d_{abc}\phi^{a}\phi^{b}\phi^{c} 
    \\
    & +\frac{1}{12}\left[  \sum_{\alpha \ \in \ \Delta_{+}} |\alpha_{a}\phi^{a}|^{3} -
\sum_{R_{f}}\sum_{\lambda_{R_{f}}} |\lambda_{i}\phi^{i} + m_{f}|^{3} \  \right].
\end{split}
\end{equation}
Here, $\alpha^{a}$ are the positive roots of the Lie algebra, and $\lambda_{R_{f}}$
are the weights of the representation $\textbf{R}_{f}$. The constants are determined
through group theory data as $c_{ab}=\Tr\{T_{a}T_{b}\}$ and $2 d_{abc} =
\Tr\{T_{a}\{T_{b},T_{c}\}\}$. It should be noted that the first line in the above
equation represents the classical part of the prepotential where-as one-loop effects
are dictated in the second line.

As the moduli space is $\mathcal{C} = \mathbb{R}^{r}/W(G)$, in the absence of matter fields,
the CB is divided into $|W(G)|$ chambers. For example, a particular chamber can be
defined as 
\begin{equation}
    \mathcal{C} = \{\phi \in \R^{r}| \expval{\phi,\alpha^{a}} >0, \ \forall
\alpha^{a} \in \Delta_{+}\},
\end{equation}
which would define a cone in the space of $\Delta_{+}$, whose boundary is
defined by $\expval{\phi,\alpha^{a}}=0$. Turning on some massless matter fields then
the moduli space receives further subdivisions. One crosses between different
chambers at a "wall" defined by $\expval{\phi,\lambda_{R}} =0$, at these loci BPS
states become massless, then at each side of the wall the quantity
$\expval{\phi,\lambda_{R}}$ has a definite sign. To study the phases of the ECB some
pictorial techniques have been establishes and are called $\it{box\ graphs}$ \cite{Hayashi:2014kca}.

So far our discussion has been suited for the weakly coupled IR description of the
theory, where a Lagrangian may be available.  However the UV-limit of the theory is
not very accessible as the coupling $g$ grows to infinity. In this limit new light
particles and strings with tension $T_{a}=\partial_{a}\mathcal{F}(\phi)$ appear in the
spectrum as their tension decreases. Moreover, from the perspective of the
renormalization group flow the existence of a UV-fixed point is not obvious. Indeed the
Lagrangian is non-renormalizable.  It turns out that we have to establish the notion
of a physical CB, defined as
\begin{equation}
    \mathcal{C}_{phy} = \{ \phi\in \mathcal{C}| \  T_{a}(\phi) > 0, \
m^{2}(\phi)>0 , \  G_{ab} > 0\}.
\end{equation}
In the case where all the masses vanish, the tensions are constrained by
$T_{a}(\phi)\geq 0$. As an example, we take $SU(2)$ gauge theory with $N_{f}$ HMs in
the fundamental $\textbf{F}$, the metric is non-degenerate and the tension is
positive, in the limit $g\rightarrow\infty$, only for $N_{f} < 8$. Thus, the
physical ECB is defined and hints at the existence of a non-trivial UV-fixed point.
The point at $N_{f}=8$ is somewhat special. Such theories are called rank one 
Seiberg theories. A special feature occurs as claimed by \cite{Seiberg:1996bd,
Morrison:1996xf, Intriligator:1997pq}, which enhances the flavour symmetry group. In
the example above, 
\begin{equation}
    G_{F}^{IR} = SO(2N_{f}) \times U_{T}(1) \ \rightarrow \ G_{F}^{UV} = E_{N_{f}+1} ,
\end{equation}
appears as a symmetry group of the super-conformal field theory (SCFT) at the
UV-fixed point. Clearly, field-theoretic analysis is not accessible in the strongly
coupled limit, and with this framework, we can not prove the claim. To gain another
perspective on this problem we turn our attention to M-theory
\cite{Intriligator:1997pq}.

\subsection{M-theory Construction of five-dimensional SCFTs}

Consider M-theory compactification on a non-compact AC Calabi-Yau 3-fold $CY_{3}$ denoted by $X$, then the low-energy theory is a 5-dimensional field theory with proprieties inherited from the 3-fold $X$, 
\begin{equation}\label{Mtheory to T}
    \text{M-theory on } \R^{1,4}\times X \ \simeq \ \mathcal{T}_{5}\{X\}.
\end{equation}
As discussed in the introduction, we get a large class of such singularities from finite subgroups of $SU(3)$ acting on $\C^{3}$. Before diving into that, let us develop some intuition in the relation between the LHS and the RHS of eq(\ref{Mtheory to T}). The proprieties of a singular $CY_{3}$ can be revealed by studying its resolutions $\widetilde{X}$. In particular, crepant resolutions and deformations are the main tools to smooth out a singular $CY_{3}$. The former are associated with the CB, while the latter are associated with the Higgs branch (HB). In this work, we focus on the crepant resolutions, {\it {\it i.e.}} we study the Coulomb branch.

A resolution $\widetilde{X}$ of $X$, $\pi : \widetilde{X} \rightarrow X$, is called a crepant resolution if $K_{\widetilde{X}} = \pi^*K_X$. Moreover, if there exists a conically singular Calabi-Yau metric on $X$ and a crepant resolution $\widetilde{X}$, then van the Coevering's theorem establishes the existence of an AC Calabi-Yau metric on $\widetilde{X}$.


Note that crepant resolutions are not unique, and a resolution corresponds to a phase in the ECB while the singular limit corresponds to the SCFT and one can interpolate between the different phases of the ECB {\it via} geometric transitions \cite{Hayashi:2014kca, Esole:2014bka}. In a complete resolution $\tilde{X}\rightarrow X$ one introduces the exceptional divisors, and the dimension of the ECB is equal to the number of exceptional divisors. Non-abelian gauge bosons arise as M2-branes wrap two-cycles $\P^{1}$'s in the compact divisor $S_{a}$\footnote{Thus, the compact divisor in this case is a $\P^{1}$ fibration over a curve $\Sigma$, $\P^{1} \hookrightarrow S_a \rightarrow \Sigma$.}, where the $\P_{a}^{1}$ satisfies the following conditions \cite{Witten:1996qb},
\begin{equation}
    S_{a}\cdot \P_{a}^{1} = -2,\ \P_{a}^{1}\cdot_{S_a} \P_{a}^{1} = 0.
\end{equation}
This type of two-cycles is therefore called a $(-2,0)$-curve in reference to its normal bundle in $\tilde{X}$. The gauge algebra $G$ is then determined from the intersections between the divisors $\{S_{a}\}$ and $\mathbb{P}_a^1$
\begin{equation}
	S_{a}\cdot \P_{b}^1 = - C_{ab},
\end{equation}
where $C_{ab}$ is the Cartan matrix of an ADE Lie algebra. M2-branes wrapping the $(-1,-1)$-curves will become hypermultiplets. A detailed discussion of the dictionary between geometry and physics is in \cite{Eckhard:2020jyr}. 

The previous discussion gives us the tools to understand the spectrum, but we also want to know how to determine the IR dynamics of the theory, namely the prepotential $\mathcal{F}_{geo}$. The prepotential in Eq. \ref{mathcalF} is at most cubic in the fields $\{\phi^{a}\}$, and this structure can be naturally obtained from triple intersection numbers of the compact and non-compact divisors\footnote{With at least one compact divisor.}. One may define the Poincar\'e dual of the K$\ddot{\text{a}}$hler class $T$ as
\begin{equation}
    T = \sum_{\mu} \varphi_{\mu} \mathcal{D}_{\mu}.
\end{equation}
Here $\{\varphi_{\mu} = (\phi_{a},g, m_{f})\}$, and $\mathcal{D}_{\mu}$ stands for both compact and non-compact divisors. The geometric prepotential is then defined as
\begin{equation}
    \mathcal{F}_{geo}(\phi^{a},g)  = - \frac{1}{6} T\cdot T\cdot T.
\end{equation}

\bigskip

The five-dimensional strings of the theory can be obtained {\it via} $M5$-branes wrapping the compact surfaces $S_{a}$. The tension of these magnetic strings are given by the K$\ddot{\text{a}}$hler volume of the compact divisors, 
\begin{equation}
    T_{a}  \sim \int_{S_{a}} \omega^{(1,1)}\wedge \omega^{(1,1)} = \text{vol}(S_{a}),
\end{equation}
which is equal to the first derivative of the prepotential with respect to $\phi^{a}$, $\partial_{a}\mathcal{F}_{geo}$. Indeed in the collapsing limit the strings become tensionless.

Thus, starting from the {\it conically} singular 3-fold $X$ that corresponds to the SCFT fixed point
one can deform along the Coulomb branch and then slide along the renormalization group flow by performing crepant resolutions. The crepant resolution is not unique, and this suggests that there are various gauge theories with a common UV-fixed point, which are called UV-dual. With this, we conclude the generic description on five-dimensional theories {\it via} M-theory, and the reader is invited to consider \cite{Xie_2017, Closset:2018bjz, Eckhard:2020jyr}. However before we turn to new theories we first recall the so-called $T_N$ theories of  \cite{Benini:2009gi} as an example of our discussion and also for future reference.\\

\subsection{$T_{N}$ theories}


The physics of M-theory on abelian orbifolds of $\C^{3}$ are perhaps the best understood examples, as these orbifolds and their resolutions are toric varieties. The example that is relevant to us is when $\Gamma =\Z_{N}\times \Z_{N}$ as the abelian group, which is described e.g. in \cite{Eckhard:2020jyr}. The two $\Z_{N}$ factors are generated by diagonal matrices with entries $(\omega,\bar{\omega},1)$ and $(1,\omega,\bar{\omega})$, where $\omega$ is an $N$-th root of unity, $\omega^N=1$. The physics of $M$ theory on $\C^{3}/\Z_{N}\times \Z_{N}$ is described by the $T_{N}$ theories, which are the five-dimensional analogue of the 4d $\N =2$ SCFTs first considered in \cite{Gaiotto:2009we, Gaiotto:2009gz}.

The three coordinate axes of the $\C^{3}/\Z_{N}\times \Z_{N}$ are three lines of $A_{N-1}$ singularities which meet at the conical singularity at the origin. 
Since codimension four $A_{N-1}$ singularities support $SU(N)$ gauge fields in $M$-theory and, that they are supported along infinite volume lines in the Calabi-Yau orbifold implies that the $T_{N}$-theories have a flavour symmetry group which is at least
\begin{equation}
   G_{F}^{UV}(T_{N}) = SU(N)\times SU(N)\times SU(N)\ ,
\end{equation}
corresponding to the three lines of $A_{N-1}$ singularities in the orbifold.


It is worth mentioning that the case $N=3$ is a Seiberg theory, {\it {\it i.e.}} a rank one theory. The flavour group in this case enhances to $E_{6}$, \cite{Benini:2009gi}. In the next section we will encounter similar phenomena, and we will leave the discussion of the enhancement to the next section.

\begin{figure}[H]
\centering
\includegraphics[scale=0.2]{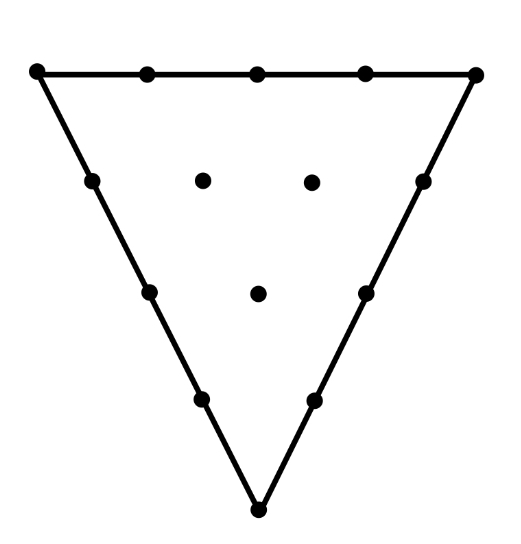}
\caption{The singular geometry of $T_{4}$. Here we choose to present the toric diagram in a hexagonal lattice therefore the $\Z_{3}$ symmetry is manifest.}
\label{N=4sing}
\end{figure}

The singular toric diagrams of the $T_{N}$ theories are interesting as they have a $\Z_{3}$ discrete symmetry. In coordinates, this is simply the cyclic permutation of the three coordinates of $\C^3$.
Therefore, we expect the $T_{N}$ theories to have such a symmetry. The symmetry permutes the three $A_{N-1}$ singularities, hence the degrees of freedom of $T_N$ theories must also form a ${\Z_3}$-invariant representation of $SU(N)^3$. However, a generic resolution of $\C^{3}/\Z_{N}\times \Z_{N}$ will break the symmetry. Nevertheless, there are $\Z_3$-invariant resolutions which can be seen manifestly in the toric description by working with a hexagonal rather than square lattice, \cite{ito1994, Ito1994CrepantRO, ItoReid, ROAN1996489}, see for example Figure \ref{N=4sing}.

\begin{figure}[H]
\centering
\includegraphics[scale=0.2]{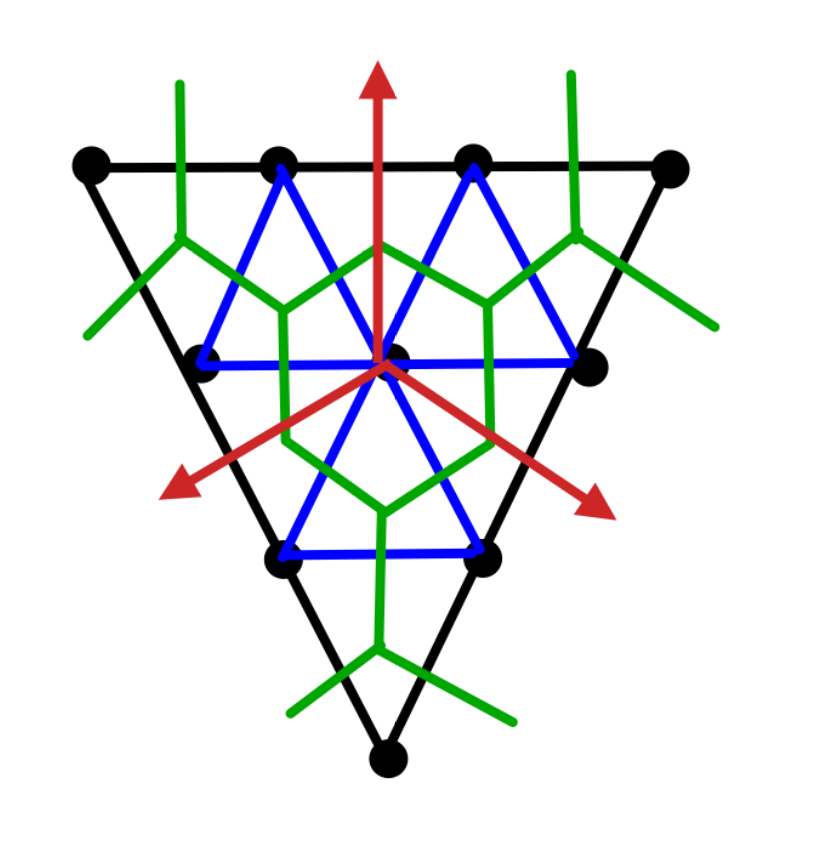}\ \ \ \ \ \ \ \ \ \ \ \  \ \ 
\includegraphics[scale=0.2]{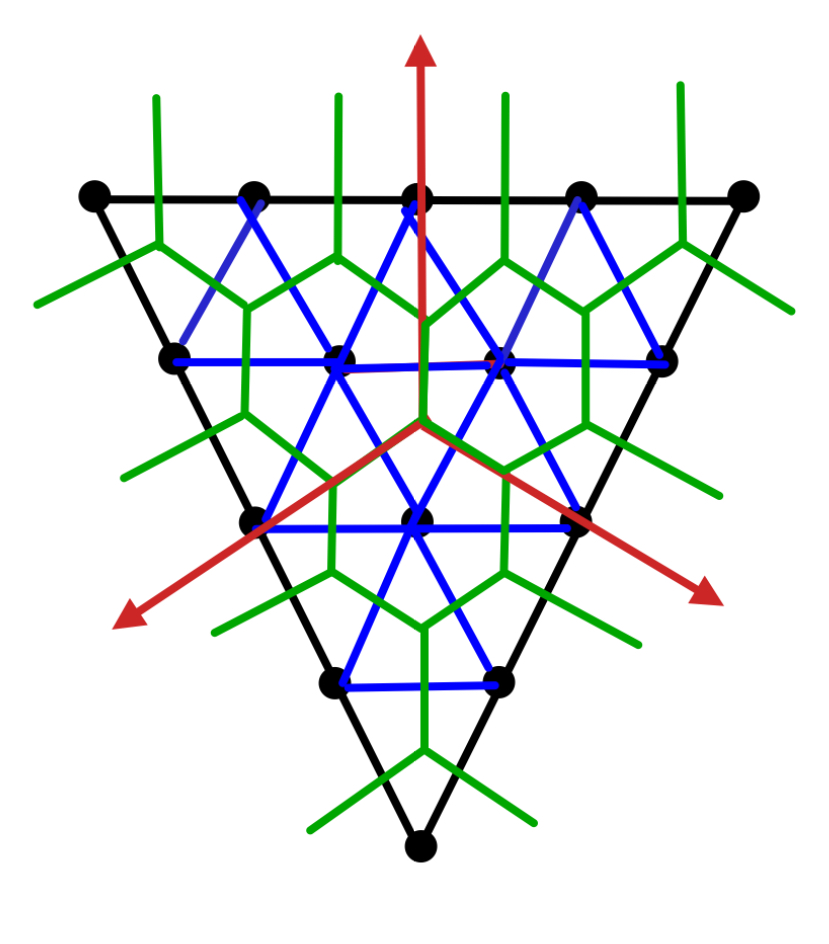}
\caption{Fully resolved toric diagrams for $N=3$ and $N=4$. The blue lines represent a resolution of the singular toric variety. The green lines correspond to the dual toric diagram identified with the 5-brane web. The $\Z_{3}$ rotational symmetry is depicted in the red arrows.}
\label{N=4 and Z3}
\end{figure}

As one moves onto the Coulomb branch, only a subset of the phases on the Coulomb branch of the $T_{N}$ theories that correspond to the $\mathbb{Z}_3$ symmetric resolutions preserves the $\Z_{3}$ symmetry, e.g. see Figure \ref{N=4 and Z3}. Therefore we will consider only the $\Z_3$-invariant Coulomb branch phases and gauge this discrete $\Z_3$ symmetry. A consistent physical interpretation is then provided as $M$-theory on the orbifold $\C^{3}/\Delta(3N^{2})$ where $\Delta(3N^2):=(\Z_{N}\times\Z_{N})\rtimes\Z_{3}$. 

As a $\Z_3$-invariant Coulomb branch phases is interpreted as a $\Z_3$-invariant crepant resolution of $\C^{3}/\Z_{N}\times \Z_{N}$, gauging the $\Z_3$-symmetry then corresponds to taking a $\Z_3$ quotient of the original resolved variety. This introduces additional codimension four $A_2$-singularities: for $N \neq 0$ mod $3$ there is one $A_2$-singularity, whilst for $N=0$ mod $3$ there are three $A_2$-singularities \cite{ItoReid, Ito1994CrepantRO, itocrepant}. The difference between the two cases is illustrated by the toric diagrams of Figure \ref{N=4 and Z3}. In the first case the fixed point set emanates from the centre of the central divisor, whereas it emanates from an intersection point in the latter case.

This implies that we will have either $SU(3)^3$ or $SU(3)$ global symmetry in the gauged theory, which, in addition implies the existence of additional degrees of freedom.

\section{The New five-dimensional SCFTs and $\Delta(3N^{2})$ group data}\label{sec:Newfive-dimensional}

In this section, we will study the physics of M-theory compactified on $\mathbb{C}^3/\Delta(3N^2)$ using the group theory data of $\Delta(3N^2)$. We will see that important physical information can be derived purely from the structure of $\Delta(3N^2)$
and the known classification of low-rank theories \cite{Jefferson:2017ahm, Bhardwaj:2018yhy} , we will identify the first few $\Delta(3N^2)$ theories with U-folds of the known and well-studied low-rank theories.

\newpage
\subsection{Group theoretical facts about $\Delta(3N^2)$ series}



For completeness, we list the finite subgroups of $SU(3)$, following the presentation in \cite{ROAN1996489}.

\begin{center}
\begin{longtable}{ | p{0.7cm} | p{1.4cm} | l | p{4cm} |} \hline

Type & Name & Generators & Comments \\ \hline

A. & Abelian & $g = \begin{pmatrix}
a & 0 & 0\\
0 & b & 0\\
0& 0 & c
\end{pmatrix} $, with $abc=1$. & Here, $\C^{3}/G$ is toric. \\ \hline

B. & & $g = \begin{pmatrix}
e & 0 & 0\\
0 & a & b\\
0& c & d
\end{pmatrix}$, with $e = \det{\begin{pmatrix}
a & b\\
c & d
\end{pmatrix}}^{-1}$ & The sub-groups $\begin{pmatrix}
a & b\\
c & d
\end{pmatrix}$ coming from $U(2)$. \\ \hline

C. & Trihedral & $H$, and
$T=\begin{pmatrix}
0 & 1 & 0\\
0 & 0 & 1\\
1& 0 & 0
\end{pmatrix}$ & Here, $H$ is an abelian sub-group. Example, $\Delta(3N^{2}) \cong (\Z_{N}\times\Z_{N}) \rtimes \Z_{3}$ \\ \hline

D. & & Type C, and $Q = \begin{pmatrix}
0 & a & 0\\
0 & 0 & b\\
c& 0 & 0
\end{pmatrix}$ & with $abc = -1$. \\ \hline

E. & $\Sigma_{108}$ & $T$, $S = \begin{pmatrix}
1 & 0 & 0\\
0 & \omega & 0\\
0& 0 & \omega^{2}
\end{pmatrix}$, $V = \frac{1}{i\sqrt{3}} \begin{pmatrix}
1 & 1 & 1\\
1 & \omega & \omega^{2}\\
1& \omega^{2} & \omega
\end{pmatrix}$& Here, $\omega = e^{2\pi i /3}$. This group is of order $108$. \\ \hline

F. & $\Sigma_{216}$ & Type E, and $UVU^{-1} = \frac{1}{i\sqrt{3}} \begin{pmatrix}
1 & 1 & \omega^{2}\\
1 & \omega & \omega\\
\omega& 1 & \omega
\end{pmatrix}$& \\ \hline

G. & $\Sigma_{648}$ & Type E, and $U = \begin{pmatrix}
\epsilon & 0 & 0\\
0 & \epsilon & 0\\
0& 0 & \epsilon\omega
\end{pmatrix}$& with $\epsilon^{3} =\omega^{2}$. \\ \hline

H. & $\Sigma_{60}$ & $T$, $E_{2} = \begin{pmatrix}
1 & 0 & 0\\
0 & -1 & 0\\
0& 0 & -1
\end{pmatrix}$, $E_{2} = \frac{1}{2}\begin{pmatrix}
-1 & \mu_{-} & \mu_{+}\\
\mu_{-} & \mu_{+} & -1\\
\mu_{+}& -1 & \mu_{-}
\end{pmatrix}$ & with $\mu_{\pm} = \frac{1}{2}(-1\pm \sqrt{5})$. \\ \hline

H$^{\ast}$. & $\Sigma_{180}$ & Type H, $W=\begin{pmatrix}
\omega & 0 & 0\\
0 & \omega & 0\\
0& 0 & \omega
\end{pmatrix}$. & Here, the matrix $W$ is the center of $SU(3)$. \\ \hline

I. & $\Sigma_{168}$ & $T$, $S_{7} = \begin{pmatrix}
\beta & 0 & 0\\
0 & \beta^{2} & 0\\
0& 0 & \beta^{4}
\end{pmatrix}$, $U = \frac{1}{\sqrt{-7}} \begin{pmatrix}
a & b& c\\
b & c & a\\
c& a & b
\end{pmatrix}$ & $\beta^{7} =1$, $a=\beta^{4}-\beta^{3}$, $b=\beta^{2}-\beta^{5}$, $c=\beta-\beta^{6}$. \\ \hline

I$^{\ast}$. & $\Sigma_{504}$ & Type I and $W$. & \\ \hline

J. & $\Sigma_{1080}$ & Type H, and $E_{4} =\begin{pmatrix}
-1 & 0 & 0\\
0 & 0 & -\omega \\
0& -\omega^{2} & 0
\end{pmatrix} $. & \\ \hline

\end{longtable}
\end{center}

We now focus on the group $\Delta(3N^2)$ which can be though of as a $\Z_{3}$ extension of the abelian finite subgroup $\Z_{N}\times \Z_{N}$ of $SU(3)$ whose generators are
\begin{equation}\label{eq:Tn_gens}
M_1 = \begin{pmatrix}
\omega & 0 & 0 \\
0 & \omega^{-1} & 0 \\
0 & 0 & 1
\end{pmatrix}, \quad
M_2 = \begin{pmatrix}
1 & 0 & 0 \\
0 & \omega & 0 \\
0 & 0 & \omega^{-1}
\end{pmatrix},
\end{equation}
where $\omega$ is an $n^{\text{th}}$ root of unity and the $\Z_{3}$ action is generated by $T$ that acts on $\mathbb{C}^3$ as
\begin{equation}\label{eq:T_gen}
T = \begin{pmatrix}
0 & 1 & 0 \\
0 & 0 & 1 \\
1 & 0 & 0
\end{pmatrix}.
\end{equation}
Immediately, we observe that $T^{-1}M_1T = M_2$ therefore equivalently we are considering the space $\mathbb{C}^3/G$ where $G = \langle M_2, T\rangle$.


\bigskip

\subsection{Five-dimensional physics from the structure of the group $\Delta(3N^2)$}

\bigskip

A general element of the $\Z_{N}\times \Z_{N}$ subgroup is diagonal and may be written as $g = \text{diag}(\omega^{a}, \omega^{b},\omega^{c}) \equiv (a,b,c)$ with $a+b+c= 0 $ mod $N$ where $\omega$ is the $N^{\text{th}}$ root of unity. The orbits of the $T$ matrix acting on the normal subgroup $H$ are the conjugacy classes, each conjugacy class consists of the elements $(a,b,c)$ and its cyclic permutation. Like-wise, the natural action of $H$ on $T$, $T^{-1}$, $gT$, and $gT^{-1}$ completes all the conjugacy classes. For any $g\in SU(3)$ one can define a function $\textit{age}(g)$ \cite{ItoReid}
\begin{equation}
\textit{age}(g) = \frac{1}{|g|}(a+b+c)\ ,
\end{equation}
where $(a,b,c)$ is obtained by diagonalizing $g$
\begin{equation}
g \xrightarrow{\text{diagonalize}} g' = \text{diag}(\omega^{a}, \omega^{b}, \omega^{c})\ .
\end{equation}
and $|g|$ is the order of $g$.
It is easy to verify that $\textit{age}(g)\in\{0,1,2\}$ for $g\in SU(3)$ and the only $g$ such that $\textit{age}(g) = 0$ is the identity. It is also obvious that all the elements in the same conjugacy class of $G\subset SU(3)$ have the same $\textit{age}(g)$. One can determine the numbers of conjugacy classes of $\Z_N \times \Z_N$ with $\textit{age}(g) =1$ or $2$ to be
\begin{equation}\label{NccHalpha=1}
\gamma_{1} = 3(N-1) + {{N - 1}\choose{2}},
\end{equation}
and
\begin{equation}\label{NccHalpha=2}
\gamma_{2} = {{N - 1}\choose{2}}.
\end{equation}


This is consistent with the formula $N^2 = 1 + \gamma_1 + \gamma_2$ which is the total number of conjugacy classes of $\Z_N \times \Z_N$. In fact, $\gamma_1$ is the total number of independent exceptional divisors of the smooth crepant resolution, whilst $\gamma_2$ is the total number of ${\it compact}$ divisors of the resolution \cite{Ito1994CrepantRO, ItoReid}. So, the number of non-compact divisors of the smooth AC Calabi-Yau threefold is $3N-3$, precisely the rank of $SU(N)^3$. Further, the rank of the five-dimensional gauge group along the Coulomb branch is $\gamma_2$. We will now extend these results to $\C^3/\Delta(3N^2)$ and its resolutions. The conjugacy classes involving the matrix $T$ all have \textit{age} equal to one.


Note, the number of non-compact four-cycles is always
\begin{equation}
rk(G^F) = |\Gamma_1| - |\Gamma_2|,
\end{equation}
and hence the rank of the flavour group.
For the case at hand, let us start with an example before discussing the general results.
Consider the geometry $\mathbb{C}^3/\Delta(27)$ where $\Delta(27)$. The conjugacy classes of $\Delta(27)$ are:
\begin{equation}
\begin{split}
\Gamma_0\ :&\ [(0,0,0)], \\
\Gamma_1\ :&\ [(0,1,2)],\ [(0,2,1)],\ [(1,1,1)],\
\\
&\ [T],\ [(0,1,2)\cdot T],\ [(0,2,1)\cdot T], \\
&\ [T^{-1}],\ [(0,1,2)\cdot T^{-1}],\ [(0,2,1)\cdot T^{-1}],
\\
\Gamma_2\ :&\ [(2,2,2)]\ .
\end{split}
\end{equation}
Therefore we have:
\begin{equation}
|\Gamma_0| = 1,\ |\Gamma_1| = 9,\ |\Gamma_2| = 1.
\end{equation}
This means there are eight non-compact divisors and one compact divisor in $\widetilde{\mathbb{C}^3/\Delta(27)}$. The gauge group of the effective theory of M-theory on $\mathbb{C}^3/\Delta(27)$ is $SU(2)$, {\it {\it i.e.}} a rank-one theory. Thus a rank-8 flavour symmetry group characterizes the SCFT fixed point. As a result, the maximal flavour group would be $E_{8}$, and this is consistent with the conjecture of Seiberg \cite{Seiberg:1996bd}.

In general, for $\C^{3}/\Delta(3N^{2})$, we have
\begin{equation}
|\Gamma_{0}| = 1,\ |\Gamma_{1}| =
\begin{cases}
&\frac{1}{3}\gamma_{1} + 2,\ N\ne 3k\\
&\frac{1}{3}(\gamma_{1}-1) + 6,\ N=3k\
\end{cases},\ |\Gamma_{2}| =
\begin{cases}
&\frac{1}{3}{{N-1}\choose{2}} \ N\ne 3k\\
&\frac{1}{3}{{N-1}\choose{2}} -\frac{1}{3}\ N=3k
\end{cases}.
\end{equation}

This result can be explained physically. In gauging the $\Z_3$ symmetry, we reduce the $SU(N)^3$ symmetry to the diagonal,
whilst the ${N-1}\choose{2}$ $U(1)$ factors on the Coulomb branch are permuted. When $N$ is not zero mod $3$, ${N-1}\choose{2}$ is an integer multiple of $3$ and all $U(1)$'s transform non-trivially under ${\mathbb Z}_3$. When $N$ is zero mod $3$, one $U(1)$ is ${\mathbb Z}_3$-invariant. Finally, the additional two or six exceptional divisors arise from the fixed points of the $\Z_3$-action and are, respectively a single or three $SU(3)$ singularities, {\it i.e.} the rank increases by two or six.

Therefore the rank of the gauge group of the five-dimensional SCFT on its Coulomb branch is
\begin{equation}\label{eq:number_g}
\text{rk}_{G}(T_{\Delta(3N^{2})}) = |\Gamma_{2}| = \begin{cases}
\frac{1}{6}\left( N^2 - 3N + 2 \right),&\ N\ne 3k\\
\frac{1}{6}\left( N^2 - 3N + 6 \right),&\ N=3k
\end{cases}\ ,
\end{equation}
and the rank of the flavour symmetry group is predicted to be
\begin{equation}\label{eq:number_f}
\text{rk}(G^F) = |\Gamma_1| - |\Gamma_2| = \begin{cases}
N+1,&\ N\ne 3k \\
N+5,&\ N=3k
\end{cases}\ .
\end{equation}

From the singularity analysis above, the flavour symmetry group $G^F$ in the two cases, therefore, contains at least $SU(N)\times SU(3)$ or $SU(N) \times SU(3)^3$. Later we will see that the latter $SU(3)^3$ is enhanced to $E_6$.

Before we conclude this section we would like to give one more example for more illustration. Let us consider M-theory compactification on $\C^{3}/\Delta(3\times 4^2)$. In this example we have
\begin{equation}
\begin{split}
\Gamma_0\ :&\ [(0,0,0)], \\
\Gamma_1\ :&\ [(0,1,3)],\ [(0,3,1)],\ [(0,2,2)],\ [(1,1,2)],
\\
&\ [T],\ [T^{-1}],
\\
\Gamma_2\ :& [(3,3,2)].
\end{split}
\end{equation}
Therefore we have:
\begin{equation}
|\Gamma_0| = 1,\ |\Gamma_1| = 6,\ |\Gamma_2| = 1 .
\end{equation}
These results can be seen partially from the toric diagram of the $T_{4}$ theory in Figure \ref{N=4 and Z3}. After the $\Z_{3}$ identification, there exists one compact divisor and four $\P^{1}$'s. The remaining $\P^{1}$'s enter {\it via} the $A_{2}$-singularity of $\Z_{3}$ acting on $\C^{3}$ as above. In total, there are five non-compact divisors; hence the corresponding five-dimensional SCFT is a rank-1 theory with five flavours. According to the known classification of rank-1 theories \cite{Seiberg:1996bd, Jefferson_2018} this is presumed to be the rank-1 $E_5$ theory, and in Section \ref{sec:T4_brane_web} we will confirm this using an analysis of a dual brane-web.

\section{$\mathbb{C}^3/\Delta(3N^2)$ as a singular Calabi-Yau hypersurface}\label{sec:HS}

In this section we will study the geometry of $\mathbb{C}^3/\Delta(3N^2)$ as a Calabi-Yau hypersurface in $\mathbb{C}^4$. We start with a $\mathbb{C}^3$ parameterized by $(x_1,x_2,x_3)$, the invariant polynomials under $M_1$, $M_2$ and $T$ in Eq. \ref{eq:Tn_gens} and \ref{eq:T_gen} are:
\begin{align}
w &= x_1^N + x_2^N + x_3^N,\nonumber \\
x &= x_1^{2N} + x_2^{2N} + x_3^{2N}, \nonumber\\
y &= x_1^Nx_2^{2N} + x_2^Nx_3^{2N} + x_3^Nx_1^{2N}, \nonumber\\
z &= x_1x_2x_3.
\end{align}
The relation $f(w,x,y,z) = 0$ between these invariant polynomials is:
\begin{equation}\label{eq:defining_eq}
-16 w^3 z^N+24 w x z^N+24 y z^N+72 z^{2N}+3 w^2 x^2-3 w^4 x-4 w^3 y+w^6+4 w x y-x^3+8 y^2 = 0.
\end{equation}
Therefore we have:
\begin{equation}
\mathbb{C}^3/G = \text{Spec}\bigg(\frac{\mathbb{C}[w,x,y,z]}{\langle f(w,x,y,z)\rangle}\bigg).
\end{equation}

One can quickly read off the codimension 2 ADE singularity from Eq. \ref{eq:defining_eq} by looking at the generators of the Jacobian ring $\mathcal{J}$ of the singular locus. There are two cases to consider. For $N\ne 3k$ the codimension 2 singularities are at:
\begin{align}
D_0 &= \{ x-w^2 = y = z = 0 \} \nonumber \\
D_1 &= \{ w^3 - 27z^N = x^3 - 27z^{2N} = y - 3z^N = 0 \}.
\end{align}
Along $D_0$ we have $\mathcal{J}_0 = \langle 1,z,z^2,\cdots, z^{N-2} \rangle$ and at $D_1$ we have $\mathcal{J}_1 = \langle 1,y \rangle$. Therefore we conclude that $G_{ADE} = SU(N)\times SU(3)$. On the other hand, for $N=3k$ the codimension 2 singularities are at:
\begin{align}
D_0 &= \{ x-w^2 = y = z = 0 \} \nonumber \\
D_k &= \{ w - 3\omega^k z^{N/3} = x - 3\omega^{2k} z^{2N/3} = y - 3z^N = 0\}\ ,
\end{align}
where $\omega$ is the cubic root of unity and $k = -1,0,1$. Along $D_0$ we have $\mathcal{J}_0 = \langle 1,z,z^2,\cdots, z^{N-2} \rangle$ and along each of $D_i$ locus we have $\mathcal{J}_k = \langle 1,y \rangle$. Therefore we conclude that $G_{ADE} = SU(N)\times SU(3)^3$. The full flavour symmetry is given by $G = U(1)^d\times G_{ADE}$ \cite{Xie_2017}. We see that $\text{rk}(G_{ADE}) = N+1$ when $N \neq 3k$ and $\text{rk}(G_{ADE}) = N+5$ for $N = 3k$ and they match the results computed {\it via} the group theoretical method in Eq. \ref{eq:number_f}. Therefore we conclude there are no extra $U(1)$ factors in the flavour algebra.

The $N=3k$ cases are particularly interesting. In this case, the hypersurface Eq. \ref{eq:defining_eq} is a degree $2N$ homogenous equation where the variables are assigned with the following weights:
\begin{align}
(y,x,w,z) = (3k, 2k, k, 1)\ ,
\end{align}
where $k = N/3$. The codimension 3 singularity of this hypersurface can be resolved {\it via} the following blow-up sequence \cite{Closset_2021}:
\begin{align}
&(y^{(3)},x^{(2)},w^{(1)},z^{(1)}|\delta_1)\nonumber \\
&(y^{(3)},x^{(2)},w^{(1)},\delta^{(1)}_i|\delta_{i+1}) \ \text{for}\ i=1,2,\cdots k-1\ ,
\end{align}
where the superscripts denote the weights of the blow-up and $\delta_i$ are sections of the exceptional divisors of the blow-up.

After applying the above sequence of blow-ups the codimension 3 singularity at $(0,0,0,0)$ is resolved though the geometry is still singular along 1-dimensional subvarieties, {\it {\it i.e.}} along codimension 2 locus in the partially resolved variety $\mathbb{C}^3/\Delta(3N^2)$. Out of the singular subvarieties, the following ones:
\begin{align}
D_i = \{ x+\frac{z^2}{2} = y = \delta_i = 0 \},\ \text{for}\ i = 1,2,\cdots, k-1
\end{align}
are compact whose Jacobian ring generators are:
\begin{align}
\mathcal{J}_i = \langle 1,z,z^2,\cdots,z^{3(k-i)-2} \rangle\ .
\end{align}
These are standard A-type singularities whose resolution leads to $|\mathcal{J}_i|$ extra compact exceptional divisors. Therefore in total we have:
\begin{equation}
r = k + \frac{1}{2}(3k-2)(k-1) = \frac{1}{6}(N^2-3N+6)\ ,
\end{equation}
which matches the result in Eq. \ref{eq:number_g}.

The compact surface $S_k:=\{\delta_k = 0\}$ is the most interesting among all the other compact 4-cycles in $\widetilde{\mathbb{C}^3/\Delta(3N^2)}$ whose defining equation is the degree 6 homogeneous equation
\begin{equation}
72 \delta_{k-1}^6-16 \delta_{k-1}^3 w^3+3 w^2 x^2-3 w^4 x-4w^3 y+w^6+24 \delta_{k-1}^3 w x+4 w x y-x^3+8 y^2+24\delta_{k-1}^3 y = 0\ ,
\end{equation}
in $\mathbb{P}^{3,2,1,1}$, and the weights of the variables are $(y,x,w,\delta_{k-1}) = (3,2,1,1)$. It is easy to show that $S_k$ is singular along the following loci:
\begin{align}
D_0 &= \{ x-w^2 = y = \delta_{k-1} = 0\},\nonumber \\
D_i &= \{ x-\frac{w^2}{3} = y - \frac{w^3}{9} = \delta_{k-1} - \frac{\omega^i}{3}w = 0 \}\ ,
\end{align}
where $i = 0,1,2$ and $\delta_0 = z$. In particular $D_0$ is at the intersection $\{\delta_{k-1} = 0\} \cap \{\delta_k = 0\}$. Therefore it is easy to see that when $k>1$ $D_0$ is the in the intersection between two compact divisors, and when $k = 1$ $D_0$ is in the intersection between the compact divisor $\{\delta_1 = 0\}$ and the non-compact divisor $\{z = 0\}$. The surface singularity along each $D_i\subset S_k$ ($i=0,1,2,3$) is an $A_2$ du Val singularity. Thus $S_k$ is a generalized del Pezzo 8 surface of the type specified by the embedding of the root system $4 A_2\hookrightarrow {E}_8$ \cite{doi:10.1142/e002, Derenthal_2013}. Therefore it is natural to assume that the theory $\mathbb{C}^3/\Delta(27)$ is a rank-1 $E_8$ theory and $\mathbb{C}^3/\Delta(3N^2)$ for $N = 3k\ (k>1)$ is a theory obtained by ``gauging'' an $A_2$ subgroup of $E_8$ flavour symmetry and couple it to the remaining part associated with the complex surface $S_1\cup S_2\cup\cdots\cup S_{k-1}$. This naturally leads to the conjecture that $E_6$ is part of the flavour symmetry in the UV limit of the $\Delta(3(3k)^2)$ theory. We note that there were earlier studies of the geometry $\mathbb{C}^3/\Delta(27)$ in \cite{Verlinde_2007, Cacciatori_2010} though the motivation was quite different. An explicit resolution of $\mathbb{C}^3/\Delta(27)$ and the intersection numbers in the compact divisor $S_1$ can be found in \cite{Cacciatori_2010}.

We emphasize that the tensor product of the codimension 2 ADE algebras of the singular CY hypersurface is not necessarily the UV flavour algebra of the five-dimensional SCFT since there might be a further enhancement in going to UV, but the rank of the flavour algebra is always preserved in the process. In other words, studying the codimension 2 ADE singularities leads to the correct number of non-compact exceptional divisors in a crepant resolution of the singular CY3 hypersurface but not necessarily the correct intersection numbers between the exceptional divisors of the resolved geometry. More refined geometrical data needs to be studied in order to determine the UV flavour algebra, and we will do this in future work.

\section{A 5-Brane Web Construction of $\Delta(3N^2)$ Theories}\label{sec:brane_web}

In this section we consider two concrete example to illustrate the difference between the $N=3k$ case and $N\neq 3k$ case, and to show that the physics of $\Delta(3N^2)$ theories can also be studied {\it via} the traditional brane web construction starting with the $(p,q)$ 5-brane web of the $T_N$ theory. Starting with the brane web of $T_N$ theory we will apply a $\mathbb{Z}_3$ action that transforms a D5-brane to an NS5-brane, an NS5-brane to a $(1,1)$ 5-brane and a $(1,1)$ 5-brane to a D5-brane. This action can be realized {\it via} $M\in SL(2,\mathbb{Z})$ acting on the $(p,q)$ charge of the 5-brane where
\begin{align}
M = \begin{pmatrix}
0 & -1 \\
1 & -1
\end{pmatrix}	\ .
\end{align} It is easy to see that $M^3 = \text{Id}_{2\times 2}$ and $M$ is equivalent to the monodromy generated by a stack of 7-branes that carries an $E_6$ algebra in the F-theory uplift up to a global $SL(2,\mathbb{Z})$ transformation. We will see how this $E_6$ algebra arises physically from the $\mathbb{Z}_3$ action on the $T_N$ brane web when $N = 3k$. We will argue that when $n\neq 3k$ this $E_6$ algebra is ``broken'' to $SU(3)$. Indeed there is another order three monodromy
\begin{align}
M = \begin{pmatrix}
-1 & 1 \\
-1 & 0
\end{pmatrix}	\ ,
\end{align}
that transforms a D5-brane to a $(1,1)$ 5-brane, a $(1,1)$ 5-brane to an NS5-brane and an NS5-brane to a D5-brane and this monodromy is equivalent to the monodromy generated by a stack of 7-branes that carries an $SU(3)$ algebra in the F-theory uplift up to a global $SL(2,\mathbb{Z})$ transformation.

The brane web techniques that we use in this section was developed and utilized in a large portion of the literature, to mention a few that we will closely follow, we direct the reader to \cite{Aharony_1997, Aharony_1998, Benini:2009gi, Kim_2015, Hayashi_2018} and the review article \cite{Giveon_1999}. We will see that the F-theory uplift play an important role in our derivation and the most relevant materials are in \cite{Gaberdiel_1998, DeWolfe_1998, Grassi_2013, Halverson_2017}.

For our purpose will need both 5-branes and 7-branes in the system. In the IIB/F-theory picture we have the brane configuration in Table \ref{tab:IIBbrane}.
\begin{table}[h]
\begin{center}
\begin{tabular}{c|c|c|c|c|c|c|c|c|c|c}
& 0 & 1 & 2 & 3 & 4 & 5 & 6 & 7 & 8 & 9 \\
\hline
$NS5$ & $\checkmark$ & $\checkmark$& $\checkmark$ & $\checkmark$ & $\checkmark$ & $\checkmark$ & & & & \\
\hline
$D5$ & $\checkmark$ & $\checkmark$& $\checkmark$ & $\checkmark$ & $\checkmark$ & & $\checkmark$ & & & \\
\hline
$(p,q)$ 5-brane & $\checkmark$ & $\checkmark$ & $\checkmark$ & $\checkmark$ & $\checkmark$ & $\circ$ & $\circ$ & & & \\
\hline
$(p,q)$ 7-brane & $\checkmark$ & $\checkmark$ & $\checkmark$ & $\checkmark$ & $\checkmark$ & & & $\checkmark$ & $\checkmark$ & $\checkmark$
\end{tabular}\caption{The branes that expand in different spacetime dimensions in the IIB/F-theory picture. The $\circ$ means that a general $(p,q)$ 5-brane lives on a line in the 56 plane.}\label{tab:IIBbrane}
\end{center}
\end{table}
We will focus on the 56-plane where all the interesting physics happen. On the 56-plane, a D5-brane ends on a D7-brane, an NS5-brane ends on a $(0,1)$ 7-brane, and a $(p,q)$ 5-brane ends on a $(p,q)$ 7-brane. To preserve supersymmetry we further require the angle $\theta$ of each $(p,q)$ 5-brane on the 56-plane satisfy $\tan(\theta) = q/p$ \cite{Aharony_1997} and we see this matches the convention we have adopted in Table \ref{tab:IIBbrane}. The 7-branes are important because they create branch cuts on the 56-plane, leading to a non-trivial deficit angle and monodromy at infinity. Moreover, introducing 7-branes provides a way to create and annihilate 5-branes when we perform Hanany-Witten moves \cite{Hanany_1997}.

\begin{figure}[h]
\begin{center}
\includegraphics[scale = 1]{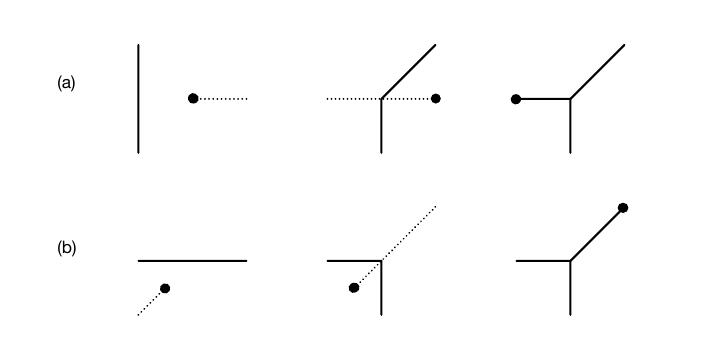}
\end{center}\caption{The equivalent configurations (a) and (b). In configuration (a) the node represents a D7-brane while in configuration (b) the nodes represents a $(1,1)$ 5-brane. The dashed lines in both configurations represent the corresponding branch cuts on the 56 plane.}
\label{fig:basic_s_rules}
\end{figure}

The (generalized) s-rule is central to this brane creation and annihilation process \cite{Hanany_1997, Gaberdiel_1998, DeWolfe_1998, Benini:2009gi}. For our purpose it is sufficient to recall that the three basic 1-junctions in configuration (a) and (b) on the 56-plane in Figure \ref{fig:basic_s_rules} are all equivalent. The last configuration of (a) and that of (b) in Figure \ref{fig:basic_s_rules} are the so-called 1-junction \cite{Benini:2009gi}. For our purpose it is also worth mentioning that a $(p,q)$ 7-brane can go across a $(p,q)$ 5-brane without creating an extra 5-brane, equivalently on can say that a $(p,q)$ 5-brane can go through the branch cut created by a $(p,q)$ 7-brane without bending.

\subsection{$\mathbb{Z}_3$ quotient of the brane web}

Before we study the examples using the brane web techniques that have been quickly reviewed, we want to comment on what we mean by a $\mathbb{Z}_3$ quotient of the brane web.

As we have mentioned before, the $\mathbb{Z}_3$ action on the brane web is realized by an order three $SL(2,\mathbb{Z})$ matrix $M$ acting on the $(p,q)$ charges of the 5-branes. Naively one would conclude that to obtain a $\mathbb{Z}_3$ quotient, the branes whose $(p,q)$ charges are on the same orbit of $M$ should be identified. But our $\mathbb{Z}_3$ action is more than simply the matrix multiplication by $M$, one also needs to specify which $(p,q)$ 5-brane is to be identified with which $(p',q')$ 5-brane when there are multiple $(p,q)$ and $(p',q')$ 5-branes in the system and obviously we require $(p',q')^t = M\cdot (p,q)^t$.

\begin{figure}[h]
\begin{center}
\includegraphics[scale = 0.7]{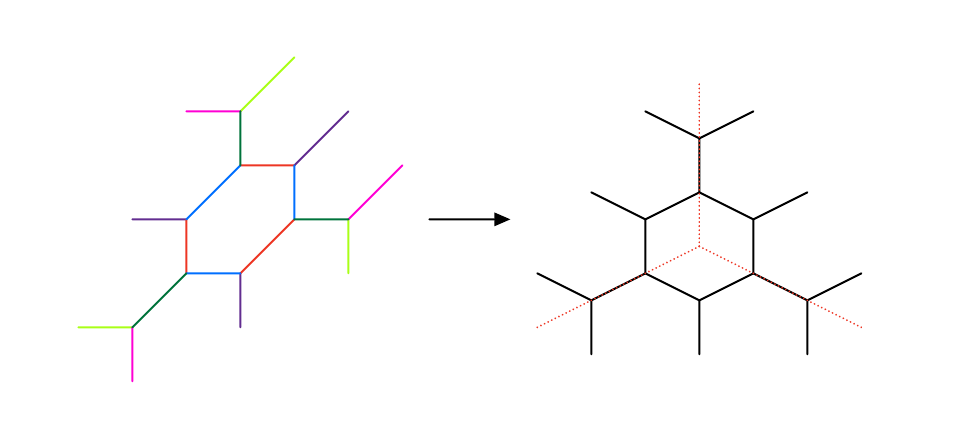}
\caption{The $\mathbb{Z}_3$ action on the $(p,q)$ 5-brane web. On the left it is a conventional $T_3$ brane web diagram that is valid at zero axion. 5-branes with the same colour are identified by the $\mathbb{Z}_3$ action. On the right we have redrawn the picture for the case where the axio-dilaton is fixed to $\tau = e^{2\pi i/3}$. We see that our $\mathbb{Z}_3$ action is realized by a $\frac{2}{3}\pi$ rotation in the tilted brane web.}
\label{fig:Z3actiononweb}
\end{center}
\end{figure}

The identification will be quite natural since we deal with a $\mathbb{Z}_3$ action on the brane web of the five-dimensional $T_N$ theory. It is best illustrated by an example shown in Figure \ref{fig:Z3actiononweb}.

At this point we remind the reader that all the brane-web diagrams with a $(p,q)$ 5-brane represented by a line segment with $\text{tan}(\theta) = q/p$ are for $\tau = i$. Physically the axio-dilaton field $\tau$ of the configuration we are interested in is not $i$, rather it restricted to be a fixed point of the $SL(2,{\mathbb Z})$ transformation $M$:
\begin{align}
\tau = \frac{-1}{\tau-1}	\qquad \Longleftrightarrow\qquad \tau = e^{2\pi i/3}\ .
\end{align}
Therefore, physically the brane-webs we are working with are tilted one as the one on the right of Figure \ref{fig:Z3actiononweb} and in that sense the $\Z_3$ action on the brane-web will indeed be a geometrical symmetry that acts as a $\frac{2}{3}\pi$ rotation in the 56-plane. But as argued in \cite{Aharony_1998} one can always ``normalize'' $\tau$ to make $\tau = i$. Therefore we will stick to the conventional way of drawing the five-dimensional brane-web diagram such that a $(p,q)$ 5-brane is represented by a line segment with $\text{tan}(\theta) = q/p$. Therefore, under this convention, the $\mathbb{Z}_3$ action can be realized by requiring that a $(p,q)$ 5-brane is represented by a line segment with length $\sqrt{p^2 + q^2}$ and putting the ``centre'' of the $T_N$ brane web at the origin of the 56-plane, then we identify a vertex at $(x,y)^t$ with a vertex at $M\cdot (x,y)^t$. We will see that the extrapolation of this action on the vertices in the brane web to an action on the whole 56-plane is useful when studying the five-dimensional Seiberg-Witten curve in Section \ref{sec:curve}.

\begin{figure}[h]
\begin{center}
\includegraphics[scale = 0.7]{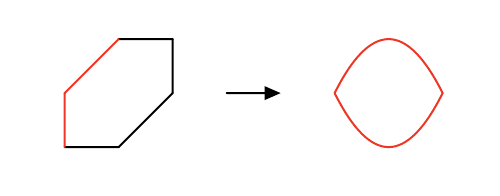}
\caption{The $\mathbb{Z}_3$ action on part of the $(p,q)$ 5-brane web. The edges in the fundamental region of the $\mathbb{Z}_3$ action is colored red.}
\label{fig:curved_web}
\end{center}
\end{figure}

We make several important remarks on the $\mathbb{Z}_3$ action on the brane web before we present the examples. Naively the identification of 5-branes due the $\mathbb{Z}_3$ action will lead to a curved brane diagram as illustrated in Figure \ref{fig:curved_web}.
We see that because of the identification, the resulting brane web which consists of the edges in the fundamental region of the $\mathbb{Z}_3$ action on the 56-plane becomes curved. This leads to a small paradox because we know that on the 56-plane a $(p,q)$ 5-brane should really be represented by a line segment whose slope is $q/p$ hence is never curved. The paradox is resolved by adding 7-branes into the system. Actually, the introduction of 7-branes is required because the $\mathbb{Z}_3$ action acts not only on the 5-branes but also extends to an action on the whole 56-plane. Hence it generates a deficit angle at infinity that corresponds to an order three monodromy on the axio-dilaton. An alternative point of view is to think of the branes as living on a cone where they do correspond to geodesics. In particular, one can plant a stack of 7-branes with an order three monodromy at the origin, and it is the branch cuts generated by the 7-branes on the 56-plane that ``bend'' the 5-branes. In particular, this bending due to the 7-brane branch cuts leads to an unconventional curved brane web. The curved brane web is a somewhat imprecise representation, and it should be understood that the bending is due to the introduction of 7-brane branch cuts. More precisely, the curved brane web lives in a space that is topologically $\mathbb{C}/\mathbb{Z}_3$ and if one would like to study the brane web in a more conventional way, {\it i.e.}, putting it back on $\mathbb{C}$, then it is necessary to introduce extra 7-branes (with their branch cuts) that encode the $\mathbb{Z}_3$ quotient structure, and in this process, the curved branes are ``straightened''.

\subsection{The $\mathbb{Z}_3$ quotient of the $T_3$ brane web}

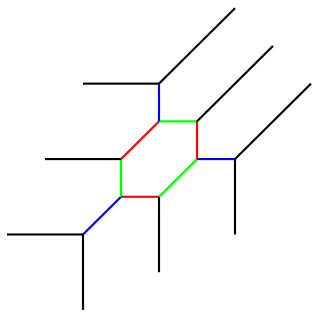
\begin{figure}[h]
\begin{center}
\begin{tikzpicture}
\draw[black, thick] (-1,-2) -- (-1,-1);
\draw[black, thick] (-2,-1) -- (-1,-1);
\draw[blue, thick] (-1,-1) -- (-1/2,-1/2);
\draw[green, thick] (-1/2,-1/2) -- (-1/2,0);
\draw[red, thick] (-1/2,-1/2) -- (0,-1/2);
\draw[black, thick] (-1/2,0) -- (-3/2,0);
\draw[red, thick] (-1/2,0) -- (0,1/2);
\draw[black, thick] (0,-1/2) -- (0,-3/2);
\draw[green, thick] (0,-1/2) -- (1/2,0);
\draw[green, thick] (0,1/2) -- (1/2,1/2);
\draw[blue, thick] (0,1/2) -- (0,1);
\draw[red, thick] (1/2,0) -- (1/2,1/2);
\draw[blue, thick] (1/2,0) -- (1,0);
\draw[black, thick] (0,1) -- (-1,1);
\draw[black, thick] (0,1) -- (1,2);
\draw[black, thick] (1,0) -- (1,-1);
\draw[black, thick] (1,0) -- (2,1);
\draw[black, thick] (1/2,1/2) -- (3/2,3/2);
\end{tikzpicture}\caption{Brane web of the $T_3$ theory. The branes labelled by the same color are to be identified upon the $\mathbb{Z}_3$ action. The identification between the out-going 5-branes is obvious hence we haven't colored them.}
\label{fig:T3_web}
\end{center}
\end{figure}

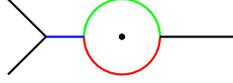
\begin{figure}[h]
\begin{center}
\begin{tikzpicture}
\draw[green, thick] (-0.5,0) arc (180:0:0.5);
\draw[red, thick] (0.5,0) arc (180:0:-0.5);
\draw[black, thick] (-1.5,-0.5) -- (-1,0);
\draw[black, thick] (-1.5,0.5) -- (-1,0);
\draw[blue, thick] (-1,0) -- (-0.5,0);
\draw[black, thick] (0.5,0) -- (1.5,0);
\filldraw[black] (0,0) circle (1pt) node[anchor=west] {};
\end{tikzpicture}\caption{A $\mathbb{Z}_3$ quotient of the brane web of the $T_3$ theory.}
\label{fig:T3_Z3_web}
\end{center}
\end{figure}

The $T_3$ brane web is given in Figure \ref{fig:T3_web}. We apply a $\mathbb{Z}_3$ action on the brane web such that the branes labelled by the same colour are identified. After the $\mathbb{Z}_3$ action, we arrive at a brane configuration illustrated in Figure \ref{fig:T3_Z3_web}. We will see that this configuration can be put into a more conventional form and is equivalent to a brane web that exhibits $E_8$ flavour symmetry.

As we have stated, the $\mathbb{Z}_3$ action acts not only on the $(p,q)$ 5-branes but also on the 56-plane where the brane web lives. In Figure \ref{fig:T3_Z3_web} the central node inside the green-red loop represents the fixed point of the $\mathbb{Z}_3$ action on the 56-plane which in particular is the fixed point of an order three action. It is then natural to conclude that the physics of the fixed point is carried by a stack of 7-branes whose world volume is along the 01234789 direction with an order three monodromy measured at infinity. In particular, we will see that this order three monodromy is realized by the stack of 7-branes that carries an $E_6$ algebra. We will adopt the convention that the $E_6$ algebra is realized by the brane system $\{\pi_1, \pi_3, \pi_1, \pi_3, \pi_1, \pi_3,\pi_1, \pi_3 \}$ \cite{Grassi_2013}. Here $\pi_1$ denotes a D7-brane and $\pi_3$ denotes a $(0,1)$ 7-brane.

\begin{figure}[h]
\begin{center}
\includegraphics[scale = 0.7]{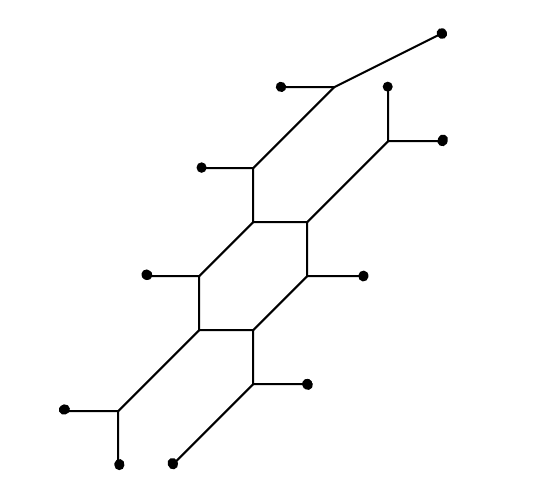}
\caption{A 5-brane web that realizes $E_8$ flavour symmetry reproduced from \cite{Kim_2015}. The nodes are the $(p,q)$ 7-branes on which a $(p,q)$ can end.}
\label{fig:E8_brane_web}
\end{center}
\end{figure}

A known 5-brane web that exhibits $E_8$ flavour symmetry (after suitable Hanany-Witten moves) was presented in \cite{Kim_2015}, and we reproduce it in Figure \ref{fig:E8_brane_web}. Note that there are other 5-brane web that realizes the $E_8$ flavour symmetry as well. The advantage of using this particular brane web is that it can be easily transformed {\it via} Hanany-Witten moves into a configuration that is very useful for our purpose.

\begin{figure}[h]
\begin{center}
\includegraphics[scale = 0.7]{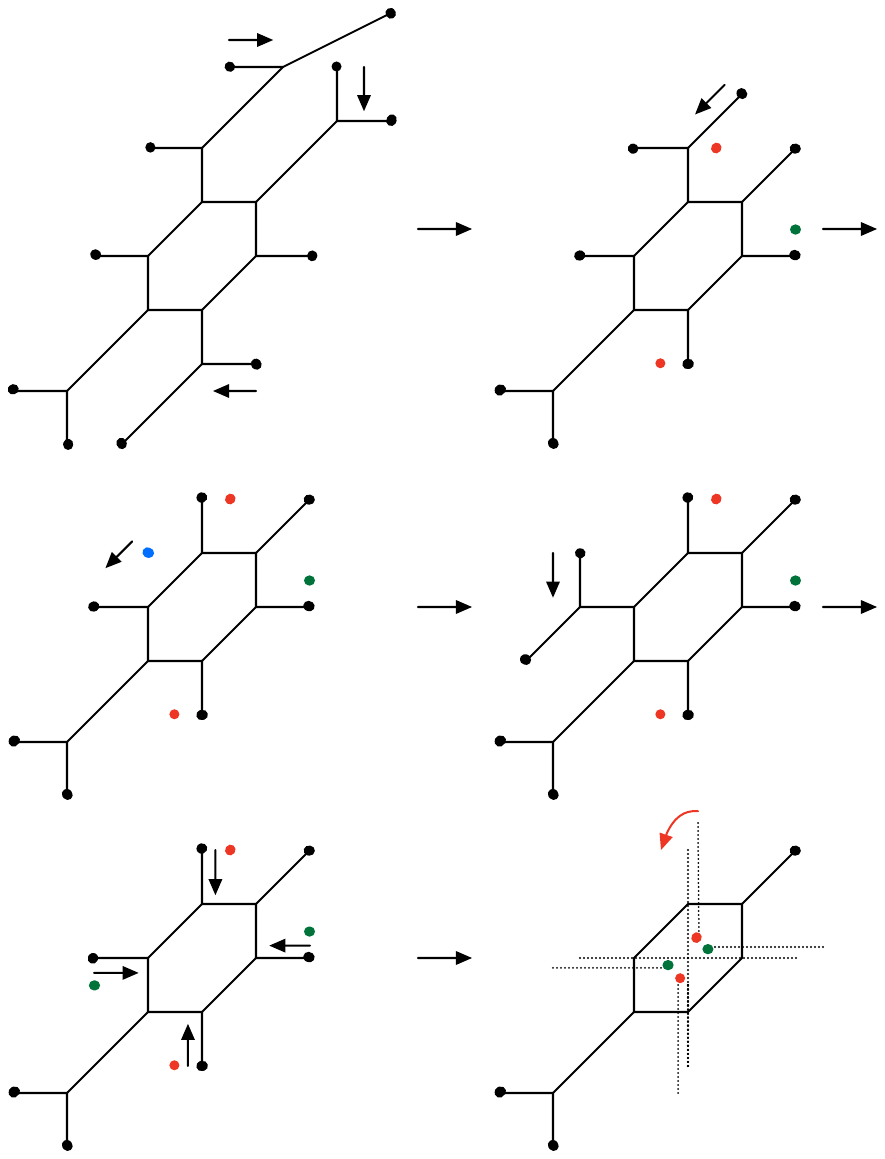}
\caption{The chain of Hanany-Witten moves. The nodes denote the 7-branes, and the black arrows denote the direction of the movement of the 7-branes. The red node denotes a D7-brane, the green node denotes a $(0,1)$ 7-brane, and the blue node denotes a $(1,1)$ 7-brane. We haven't coloured the 7-branes that are connected with a 5-brane since their charges are determined by the 5-branes that end on them, whose charges are in turn determined by their angle $\theta$ on the 56-plane. The red arrow in the last step denotes the starting branch cut and the direction that we apply to compute the total monodromy at infinity.}
\label{fig:HW_moves_E8}
\end{center}
\end{figure}

We then perform the Hanany-Witten moves as illustrated in Figure \ref{fig:HW_moves_E8}.
At the last step we have moved all the 7-branes inside the central square. Now it is easily seen that under our convention the total monodromy created by all the eight branch cuts is $(M_{\pi_1}\cdot M_{\pi_3})^4$ where
\begin{align}
M_{\pi_1} = \begin{pmatrix}
1 & 1 \\
0 & 1
\end{pmatrix},\ M_{\pi_3} = \begin{pmatrix}
1 & 0 \\
-1 & 1
\end{pmatrix}.
\end{align}
The above monodromy is exactly the same monodromy generated by the stack of 7-branes $\{\pi_1, \pi_3, \pi_1, \pi_3, \pi_1, \pi_3,\pi_1, \pi_3 \}$ at infinity as anticipated before. In particular we have
\begin{equation}
M_{\text{total}} = (M_{\pi_1}\cdot M_{\pi_3})^4 = \begin{pmatrix}
0 & -1 \\
1 & -1
\end{pmatrix}
\end{equation}
and $M_{\text{total}}^3 = \text{Id}_{2\times 2}$.

It is then natural to conclude that the last brane configuration of the chain of HW moves shown in Figure \ref{fig:HW_moves_E8} is indeed equivalent to the brane configuration in Figure \ref{fig:T3_Z3_web}. The curved 5-brane configuration in Figure \ref{fig:T3_Z3_web} can be put into the more conventional form of a brane web configuration whose basic building blocks are 1-junctions and the 7-branes with their corresponding branch cuts. Using the above HW moves, we see that the brane configuration associated with the $\mathbb{Z}_3$ quotient of the $T_3$ theory is indeed equivalent to a brane web that exhibits $E_8$ flavour symmetry, which further confirms that in the UV limit M-theory compactified on $\mathbb{C}^3/\Delta(27)$ is the rank-1 $E_8$ theory.

\subsection{The $\mathbb{Z}_3$ quotient of the $T_4$ brane web}\label{sec:T4_brane_web}

\begin{figure}[h]
\begin{center}
\includegraphics[scale = 0.7]{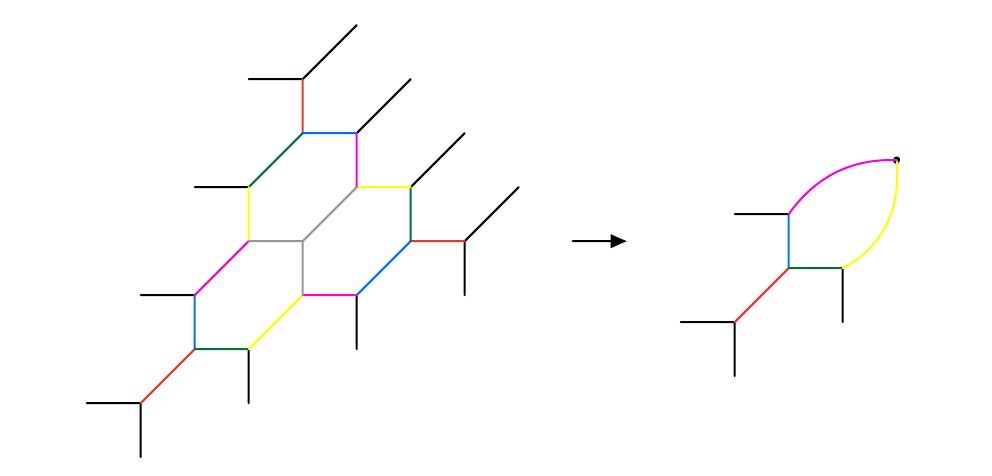}
\caption{The brane web of the $T_4$ theory and its $\mathbb{Z}_3$ quotient. The branes labelled by the same color are identified by the $\mathbb{Z}_3$ action. The node at the top-right in the $T_4/\mathbb{Z}_3$ brane configuration denotes the fixed point in the 56-plane under the $\mathbb{Z}_3$ action.}
\label{fig:T4andT4_Z3}
\end{center}
\end{figure}

Now we consider the $\mathbb{Z}_3$ quotient of the brane web of $T_4$ theory as shown in Figure \ref{fig:T4andT4_Z3}. Here again we obtain a curved brane configuration and we will see again it can be put into the more conventional form.
Note that in the $\mathbb{Z}_3$ quotient brane configuration on the left of Figure \ref{fig:T4andT4_Z3} the gray edge in the center of the original $T_4$ brane web has been ``contracted'' to the fixed point. We will argue that the effect of this contraction to the fixed point is to break the $E_6$ flavour symmetry carried by the stack of 7-branes at the fixed point to an $SU(3)$ symmetry. In particular, after lift to F-theory, this $SU(3)$ flavour symmetry is realized by a type $IV$ fibration whose monodromy is again order three.

\begin{figure}[h]
\begin{center}
\includegraphics[scale = 0.7]{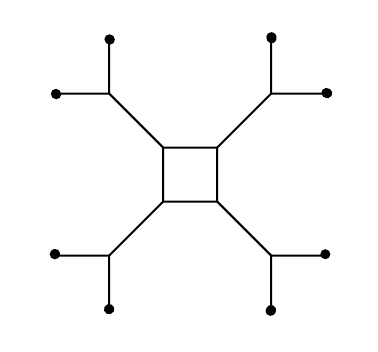}
\caption{A 5-brane web that realizes $E_5$ flavour symmetry reproduced from \cite{Kim_2015}. The nodes are the $(p,q)$ 7-branes on which a $(p,q)$ brane can end.}
\label{fig:E5_brane_web}
\end{center}
\end{figure}

\begin{figure}[h]
\begin{center}
\includegraphics[scale = 0.7]{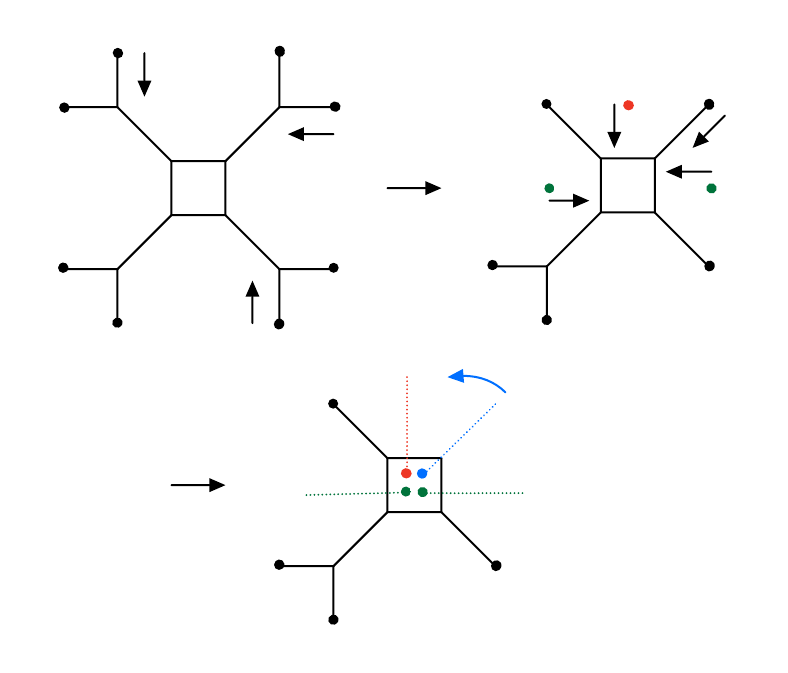}
\caption{The chain of Hanany-Witten moves. The nodes denote the 7-branes, and the black arrows denote the direction of the movement of the 7-branes. The red node denotes a D7-brane, the green node denotes a $(0,1)$ 7-brane, and the blue node denotes a $(1,1)$ 7-brane. The blue arrow in the last step denotes the starting branch cut and the direction that we apply to compute the total monodromy at infinity.}
\label{fig:HW_moves_E5}
\end{center}
\end{figure}

In this case we start with the brane web of the rank-1 $E_5$ theory as shown in Figure \ref{fig:E5_brane_web} which we reproduce from \cite{Kim_2015}.
Again we apply a chain of HW moves as shown in Figure \ref{fig:HW_moves_E5}.
Again at the last step, we have moved all the 7-branes inside the central square, and we can move these 7-branes arbitrarily close to the top-right corner of the central square, which results in a brane configuration that is equivalent to the curved brane configuration on the right of Figure \ref{fig:T4andT4_Z3} which is the $\mathbb{Z}_3$ quotient of the $T_4$ brane web. In particular, in this case, the monodromy generated by the branch cuts at infinity is
\begin{equation}
M_{\text{total}} = M_{\pi_2}\cdot M_{\pi_1}\cdot M_{\pi_3}\cdot M_{\pi_3} = \begin{pmatrix}
-2 & 1 \\
-3 & 1
\end{pmatrix}\ ,
\end{equation}
where $M_{\pi_2} = \begin{pmatrix}
0 & 1 \\
-1 & 2
\end{pmatrix}$ and it is immediately seen that $M_{\text{total}}^3 = \text{Id}_{2\times 2}$. It is also easy to check {\it via} string junction computations \cite{Grassi_2013} that the 7-brane system $\{\pi_2, \pi_1, \pi_3, \pi_3 \}$ carries an $SU(3)$ algebra.

Therefore we conclude that the $\mathbb{Z}_3$ quotient of the $T_4$ brane web is equivalent to the brane web of rank-1 $E_5$ theory up to HW moves, and this matches our previous results that M-theory compactified on $\mathbb{C}^3/\Delta(48)$ leads to a rank-1 theory with a rank-5 flavour symmetry.

\bigskip

The two concrete examples we provide in this section illustrate how our construction of the $\mathbb{Z}_3$ quotient of the brane web of $T_N$ theory can be put into a more conventional form using 1-junctions and 7-branes as fundamental building blocks {\it via} HW moves which has been studied in detail in the literature. In particular, we see that when $N=3k$ the fixed point on the 56-plane is not on the brane web, and it exhibits a monodromy at infinity generated by a stack of 7-branes that carries an $E_6$ algebra. On the other hand, when $N\neq 3k$ the fixed point is on the brane web and the monodromy at infinity is ``broken'' to that of a stack of 7-branes that carries an $SU(3)$ algebra. In the F-theory uplift, the $E_6$ algebra is realized by a type $IV^*$ fibration, and the $SU(3)$ algebra is realized by a type $IV$ fibration, and the monodromies of these two types of fibrations are both order three \cite{Halverson_2017}.

\section{$\Z_3$ automorphism on the five-dimensional Seiberg-Witten curve of the $T_N$
theory}\label{sec:curve}

We have already seen that in the brane web picture, there is a $\mathbb{Z}_3$ action
on the 56-plane, which we parameterize by $(s,t)$. At the $\mathbb{Z}_3$-symmetric
value of $\tau=\omega=e^{2i\pi/3}$ D5-branes and NS5-branes no longer intersect at
right angles, as depicted in the right-hand side of Figure \ref{fig:Z3actiononweb}.
However we can choose $s$ and $t$ to be non-orthogonal:
\begin{align}
s = \frac{\sqrt{3}}{2} x^5 + \frac12 x^6 \qquad t = x^6\ ,
\end{align} so that the metric in the 56-plane is
\begin{align}\label{stmetric}
ds^2_{56} =\frac43(ds^2+dt^2-dsdt) \ .
\end{align}
In these coordinates the slope of a $(p,q)$ 5-brane satisfies
\begin{equation}\label{eq:pq_condition_1}
\Delta s:\Delta t = p:q\ ,
\end{equation}
{\it {\it i.e.}} D5-branes sit at constant values of $t$, $NS5$-branes at constant values
of $s$ and $(1,1)$ 5-branes at constant values of $s-t$.

To construct the five-dimensional Seiberg-Witten curve we compactify the $x^4=s_t$ direction,
which we can then use to T-dualise to type IIA. Here we introduce an M-theory
circle direction with coordinate $t_t$. Thus $(s_t,t_t)$
parameterize an extra torus $T^2$ and we define the coordinates on
$\mathbb{R}^2\times T^2$ to be:
\begin{equation}
\begin{split}
& \tilde{s} = s + is_t \\
& \tilde{t} = t + it_t\ .
\end{split}
\end{equation}
On $\mathbb{R}^2\times T^2$ the slope condition Eq. \ref{eq:pq_condition_1} becomes
\cite{Kol:1997fv,Brandhuber_1997, Aharony_1997,Aharony_1998}
\begin{equation}\label{eq:pq_condition_2}
\Delta\tilde{s}:\Delta\tilde{t} = p:q\ .
\end{equation}
It is interesting to see how {\it via} the chain of dualities one can uplift the
$\mathbb{Z}_3$ action that has been described in Section \ref{sec:brane_web} to a
M5-brane configuration defined by a complex curve $C\subset\mathbb{R}^2\times T^2$,
or equivalently its image under an exponentiating map to
$\mathbb{C}^*\times\mathbb{C}^*$. This is the Seiberg-Witten curve of the
five-dimensional theory $T_{\Delta(3N^{2})}$ previously described, here compactified
on an extra circle as required by the duality chain to M-theory
\cite{Benini:2009gi}. We will mainly follow the conventions in \cite{Aharony_1998}.

Recall that on the 56-plane the $\mathbb{Z}_3$ action acts on the $(p,q)$ charge as
matrix multiplication $M\cdot(p,q)^t$. We consider the ray from the origin defined
by a pair of relative prime numbers $(p,q)$. Clearly the orbit of the $\mathbb{Z}_3$
action is
\begin{equation}
(p,q)\rightarrow M\cdot (p,q)\rightarrow M^2\cdot(p,q)\ .
\end{equation}
We could extrapolate the $\mathbb{Z}_3$ action on all pairs of relatively prime
numbers $(p,q)$ to the whole 56-plane. Therefore for any point $(s,t)$ on the
56-plane the orbit of $M$ is:
\begin{equation}
(s,t)\rightarrow M\cdot(s,t)\rightarrow M^2\cdot(s,t)\ .
\end{equation}
We could further extend this action to $\mathbb{R}^2\times T^2$ therefore, we have:
\begin{equation}
(\tilde{s},\tilde{t})\rightarrow M\cdot(\tilde{s},\tilde{t})\rightarrow
M^2\cdot(\tilde{s},\tilde{t})\ .
\end{equation}
Therefore on $\mathbb{R}^2\times T^2$ the $\mathbb{Z}_3$ is generated by
\begin{equation}
\tilde{s}\rightarrow -\tilde{t},\ \tilde{t}\rightarrow \tilde{s} - \tilde{t}\ .
\end{equation}

Next we must check that this action preserves supersymmetry. This will be the case
if it preserves the complex structure and associated K$\ddot{\text{a}}$hler metric. If we introduce
\begin{align}
z_1 = \tilde s+\omega \tilde t\qquad z_2 = \tilde s+\bar \omega\tilde t\ ,
\end{align}
where $\omega = e^{2\pi i/3}$ then we see that $dz_1\wedge dz_2$ is indeed invariant
and furthermore the $\ddot{\text{a}}$ metric
\begin{align}
ds^2 = \frac{2}{3}(dz_1\otimes d\bar z_1 +dz_2\otimes d\bar z_2 )\ ,
\end{align}
agrees with (\ref{stmetric}) on the $\tilde s= s, t=\tilde t$ plane.

Since $\tilde{s}$ and $\tilde{t}$ are not single-valued we introduce
\begin{equation}\label{eq:define_xy}
x = \exp (\tilde{s}),\ y = \exp (\tilde{t})\ ,
\end{equation}
and the $\mathbb{Z}_3$ action on $(x,y)\in\mathbb{C}^*\times\mathbb{C}^*$ becomes
\begin{equation}\label{eq:r_action}
(x, y)\rightarrow (\frac{1}{y}, \frac{x}{y})\ .
\end{equation}

Obviously the only fixed point of the $\mathbb{Z}_3$ action on 56-plane is the
origin $(s,t) = (0,0)$ therefore on $\mathbb{R}^2\times T^2$ we would consider the
points of the form
\begin{equation}\label{eq:fixed_pts_1}
(\tilde{s}, \tilde{t}) = (0+is_t, 0+it_t)\ ,
\end{equation}
as the candidates of fixed points of the $\mathbb{Z}_3$ action extended to
$\mathbb{R}^2\times T^2$. More concretely we will consider the fixed points of the
action
\begin{equation}
s_t\rightarrow -t_t,\ t_t\rightarrow s_t - t_t\ .
\end{equation}
Recall that $(s_t, t_t)$ parameterize a $T^2$, therefore they are periodic variables
and we could define the periodicity to be $2\pi$ such that no extra normalization is
needed in defining $(x,y)$ variables as in eq(\ref{eq:define_xy}). It is then easy to
see that the introduction of the extra $T^2$ leads to three fixed point whose
coordinates on $T^2$ are:
\begin{equation}
q_{1} = (0,0),\ q_2 = (\frac{2\pi}{3}, \frac{4\pi}{3}),\ q_3 = (\frac{4\pi}{3},
\frac{2\pi}{3})\ .
\end{equation}
Plugging the above coordinates into eq(\ref{eq:fixed_pts_1}) and exponentiating the
variables as in eq(\ref{eq:define_xy}) we see the fixed points are:
\begin{equation}
p_1 = (1,1),\ p_2 = (\omega, \omega^2),\ p_3 = (\omega^2, \omega)\ ,
\end{equation}
where $\omega$ is the cube root of unity. We see that in some sense the fixed points
on $\mathbb{R}^2\times T^2$ are all ``descendants'' of the only fixed point $(0,0)$
on the 56-plane under the $\mathbb{Z}_3$ action.

\begin{figure}[h]
\begin{center}
\begin{tikzpicture}
\filldraw[black] (0,0) circle (1pt) node[anchor=east] {$a_1$};
\filldraw[black] (0,1) circle (1pt) node[anchor=east] {$a_2$};
\filldraw[black] (0,2) circle (1pt) node[anchor=east] {$a_3$};
\filldraw[black] (0,3) circle (1pt) node[anchor=east] {$a_4$};
\filldraw[black] (0,4) circle (1pt) node[anchor=east] {$a_5$};
\filldraw[black] (0,5) circle (1pt) node[anchor=west] {};
\filldraw[black] (1,0) circle (1pt) node[anchor=west] {};
\filldraw[black] (1,1) circle (1pt) node[anchor=south west] {$u_1$};
\filldraw[black] (1,2) circle (1pt) node[anchor=south west] {$u_2$};
\filldraw[black] (1,3) circle (1pt) node[anchor=west] {};
\filldraw[black] (1,4) circle (1pt) node[anchor=west] {};
\filldraw[black] (2,0) circle (1pt) node[anchor=west] {};
\filldraw[black] (2,1) circle (1pt) node[anchor=west] {};
\filldraw[black] (2,2) circle (1pt) node[anchor=west] {};
\filldraw[black] (2,3) circle (1pt) node[anchor=west] {};
\filldraw[black] (3,0) circle (1pt) node[anchor=west] {};
\filldraw[black] (3,1) circle (1pt) node[anchor=west] {};
\filldraw[black] (3,2) circle (1pt) node[anchor=west] {};
\filldraw[black] (4,0) circle (1pt) node[anchor=west] {};
\filldraw[black] (4,1) circle (1pt) node[anchor=west] {};
\filldraw[black] (5,0) circle (1pt) node[anchor=west] {};
\draw[gray, thick] (0,0) -- (1,1);
\draw[gray, thick] (0,0) -- (0,5);
\draw[gray, thick] (0,5) -- (5,0);
\draw[gray, thick] (5,0) -- (0,0);
\draw[red, thick] (0,0) -- (0,5);
\draw[red, thick] (0,5) -- (1,3);
\draw[red, thick] (1,3) -- (1,1);
\draw[red, thick] (1,1) -- (0,0);
\end{tikzpicture}\caption{The toric diagram of
$\mathbb{C}^3/(\mathbb{Z}_5\times\mathbb{Z}_5)$. The red lines illustrate the
fundamental domain of nodes after the identification under the $\mathbb{Z}_3$
action.}\label{fig:T5_symm_res}
\end{center}
\end{figure}
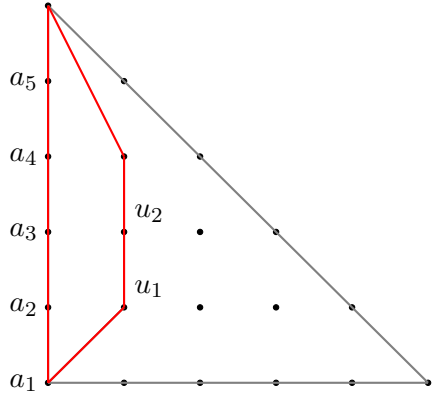

We illustrate the idea using two examples $N = 3$ and $N = 5$. The toric diagram of
$\mathbb{C}^3/\mathbb{Z}_3\times\mathbb{Z}_3$ in the ordinary square lattice is
shown in Figure \ref{fig:T3_symm_res}, and a toric diagram of
$\mathbb{C}^3/\mathbb{Z}_5\times\mathbb{Z}_5$ is shown in Figure
\ref{fig:T5_symm_res}. We will identify the nodes under the $\mathbb{Z}_3$ action in
the manner presented in \cite{doi:10.1142/S0129167X95000043, ito2001mckay}.
\begin{figure}[h]
\begin{center}
\begin{tikzpicture}
\filldraw[black] (0,0) circle (1pt) node[anchor=east] {$a_1$};
\filldraw[black] (0,1) circle (1pt) node[anchor=east] {$a_2$};
\filldraw[black] (0,2) circle (1pt) node[anchor=east] {$a_3$};
\filldraw[black] (0,3) circle (1pt) node[anchor=west] {};
\filldraw[black] (1,0) circle (1pt) node[anchor=west] {};
\filldraw[black] (1,1) circle (1pt) node[anchor=south west] {$u_1$};
\filldraw[black] (1,2) circle (1pt) node[anchor=west] {};
\filldraw[black] (2,0) circle (1pt) node[anchor=west] {};
\filldraw[black] (2,1) circle (1pt) node[anchor=west] {};
\filldraw[black] (3,0) circle (1pt) node[anchor=west] {};
\draw[gray, thick] (0,3) -- (3,0);
\draw[gray, thick] (3,0) -- (0,0);
\draw[red, thick] (0,0) -- (0,3);
\draw[red, thick] (0,3) -- (1,1);
\draw[red, thick] (1,1) -- (0,0);
\end{tikzpicture}\caption{The toric diagram of
$\mathbb{C}^3/(\mathbb{Z}_3\times\mathbb{Z}_3)$. The red lines illustrate the
fundamental domain of nodes after the identification under the $\mathbb{Z}_3$
action.}\label{fig:T3_symm_res}
\end{center}
\end{figure}
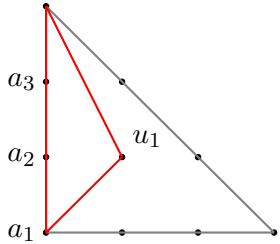

The (non-generic) Riemann surface $C_{T_3}\subset\mathbb{C}^*\times\mathbb{C}^*$
given by the Newton polytope shown in Figure \ref{fig:T3_symm_res} is an algebraic
curve given by:
\begin{equation}
f_{T_3}(x,y) = a_1(x^3+y^3+1) + a_2(x^2+x y^2+y)+ a_3 (x^2 y+x+y^2) + u x y\ ,
\end{equation}
while the (non-generic) Riemann surface
$C_{T_5}\subset\mathbb{C}^*\times\mathbb{C}^*$ given by the Newton polytope shown in
Figure \ref{fig:T5_symm_res} is an algebraic curve given by:
\begin{align}
\begin{split}
f_{T_5}(x,y) =&\ a_1 (x^5+y^5+1) + a_2(x^4+x y^4+y) + a_3(x^2 y^3+ x^3+ y^2)+
a_4(x^3 y^2+ x^2+ y^3) \\
&\ +a_5(x^4 y+ x+ y^4) + u_1 (x^3 y+xy^3+x y) + u_2 (x^2 y^2+x^2 y+x y^2)\ .
\end{split}
\end{align}
Both curves admit an automorphism $r:x\rightarrow\frac{1}{y},\
y\rightarrow\frac{x}{y}$ and it is easy to see that $r^3 = \text{Id}$. On
$\mathbb{C}^*\times\mathbb{C}^*$ the fixed points are:
\begin{align}
p_1 = (1,1),\ p_2 = (\omega, \omega^2),\ p_3 = (\omega^2, \omega)\ ,
\end{align}
where again $\omega$ is the cube root of unity. It is easy to check that for a
generic choice of the coefficients $a_i$ none of the above fixed points is on
$C_{T_3}$ while the $p_2$ and $p_3$ are on $C_{T_5}$. We denote the $r$-quotient of
$C_{T_3}$ and $C_{T_5}$ by $\mathcal{C}_{T_3}$ and $\mathcal{C}_{T_5}$ respectively,
applying Riemann-Hurwitz formula we have:
\begin{align}
& 2g(C_{T_3}) - 2 = 3(2g(\mathcal{C}_{T_3}) - 2),\nonumber \\
& 2g(C_{T_5}) - 2 = 3\left(2g(\mathcal{C}_{T_5}) - 2 + \frac{4}{3}\right).
\end{align}
It is easy to check that $g(C_{T_3}) = 1$ and $g(C_{T_5}) = 6$ therefore we have
$g(\mathcal{C}_{T_3}) = 1$ and $g(\mathcal{C}_{T_5}) = 2$.

In general, for any $N,i,j$ such that $N-i-j\geq 0$, we can tune the generic
algebraic curve $C$ associate with the Newton polygon to make the following three
terms have the same coefficient:
\begin{align}
x^{N-i-j}y^i,\ x^jy^{N-i-j},\ x^iy^j\ ,
\end{align}
such that the sum of the above three monomials $\sigma_{i,j}$ is invariant under
$r$. It is also easy to check that for $N=3k$, $\sigma_{i,j}\neq 0$ when evaluated
at $p_1$, $p_2$ or $p_3$ while for $N\neq 3k$, $\sigma_{i,j}(p_1) \neq 0$ and
$\sigma_{i,j}(p_2) = \sigma_{i,j}(p_3) = 0$. In particular, the curve
$C_{T_N}:=\sum_{i,j}C_{i,j}\sigma_{i,j}$ is smooth therefore its genus is given by
the genus-degree formula $g(C_{T_N}) = \frac{1}{2}(N-1)(N-2)$.

Therefore, with the order three automorphism $r$ on $C_{T_N}$, for $N = 3k$ we have:
\begin{equation}
2g(C_{T_N}) - 2 = 3(2g(\mathcal{C}_{T_N}) - 2)\ ,
\end{equation}
therefore
\begin{equation}
g(\mathcal{C}_{T_N}) = \frac{1}{6}N^2 - \frac{1}{2}N + 1\ ,
\end{equation}
while for $N \neq 3k$ we have:
\begin{equation}
2g(C_{T_N}) - 2 = 3\left(2g(\mathcal{C}_{T_N}) - 2 + \frac{4}{3}\right)\ ,
\end{equation}
therefore
\begin{equation}
g(\mathcal{C}_{T_N}) = \frac{1}{6}N^2 - \frac{1}{2}N + \frac{1}{3}.
\end{equation}
The genre of the curves count the rank of the corresponding gauge theories and the
above results match the results given in Eq. \ref{eq:number_g}.

In addition to the Seiberg-Witten curve $f(x,y)=0$ we can construct the Seiberg-Witten differential from $\lambda_{SW}=  \tilde s d\tilde t$. This will take the  same form as the Seiberg-Witten curve for the $T_N$ theory, as constructed in \cite{Benini:2009gi}, but restricted to the set of moduli that preserve the invariance under ${\mathbb Z}_3$. However for  $N\ne 3k$ we also find two additional singular points   arise and these could require more care.

\subsection{Five-dimensional Seiberg-Witten curve from projective Geometric Invariant Theory}

It is important to look carefully not only at the geometry of the Seiberg-Witten curve $\mathcal{C}$, but also at its embedding into a complex 
surface $X$ since both give important physical data of the M5-brane theory. Indeed, the M5-brane configuration is determined by the pair $(X, \mathcal{C})$ where $X$ is a symplectic complex surface and $\mathcal{C}\subset X$ is a complex curve \cite{Witten_1997}. In the spirit of \cite{Witten_1997} the gauge symmetry of the  M5-brane theory is determined by $\mathcal{C}$ while the flavour symmetry is determined by both $X$ and $\mathcal{C}$. In our cases, the embedded surface will be denoted by $X=S$ with the embedded curve denoted $\mathcal{C}=C^G$. For computational purposes, to account for the $\mathbb{Z}_3$-action, it is convenient to embed the story in a weighted projective space so that $C^G \subset S \subset w\mathbb{P}$. For this purpose, we will see that it is convenient to consider the compactification of the curve $C\subset \mathbb{C}^*\times\mathbb{C}^*$ into $\mathbb{P}^2$. For the background of the computations in this subsection, one can check the notes \cite{MelvinHochster, VictoriaHoskins, hosgood2020introduction} and the book \cite{mumford2013red}.

While the $r$-action assumes a non-linear form \ref{eq:r_action} on $\mathbb{C}^*\times\mathbb{C}^*$ we see that in $\mathbb{P}^2$ the uplift of $r$-action takes a simpler form which is the matrix multiplication
\begin{align}
M_r \left(= T \right) = \begin{pmatrix}
0 & 1 & 0 \\
0 & 0 & 1 \\
1 & 0 & 0
\end{pmatrix}\ ,
\end{align}
on the homogeneous coordinates $[x:y:z]$ of $\mathbb{P}^2$. Most generally we could consider a subvariety $V = V(I) \subset \text{Spec}(R) $ where the ideal $I$ is generated by a set of polynomials $\langle f\rangle$ in the coordinate ring $R$ of the ambient space. The homogeneous coordinate ring of a projective subvariety $V\subset\mathbb{P}^n$ is defined as:
\begin{equation}
R:=R(V) = R(\mathbb{P}^N)/I(V)\ ,
\end{equation}
and the following relation holds \cite{VictoriaHoskins}:
\begin{equation}
R = A(\mathbb{C}^{N+1})/I(\widetilde{V}) = A(\widetilde{V})\ ,
\end{equation}
where $A(-)$ is the affine coordinate ring of an affine variety and $\widetilde{V}$ is the affine cone over $V$ defined as:
\begin{equation}
\widetilde{V} = \{0\}\cup\{(x_0,x_1,\cdots,x_N)\in\mathbb{C}^{N+1}-\{0\}\ |\ [x_0:x_1:\cdots:x_N]\in V\subset\mathbb{P}^N\}\ .
\end{equation}
We then consider the quotient $R^G:=A(\widetilde{V})^G$ where $G$ is a reductive algebraic group. We require the $G$-action be linear, {\it {\it i.e.}}, its action lifts from $\mathbb{P}^N$ to its affine cone $\mathbb{C}^{N+1}$. Since $G$ preserves the grading on $R$, $R^G$ will be a graded subalgebra of $R$ hence we can construct the projective variety
\begin{equation}
V^G := \text{Proj}(R^G)\ ,
\end{equation}
where $\text{Proj(-)}$ can be constructed as follows.

Note that if a finite set of homogeneous generators of a ring ${\cal R}$ defines a surjective map
\begin{equation}\label{eq:relations_invpolys}
\mathbb{C}[x_0,x_1\cdots,x_N]\rightarrow {\cal R},
\end{equation}
whose kernel is an ideal ${\cal I}$, then $\text{Proj}({\cal R}) = V({\cal I})\subset \mathbb{P}^N$. Note that the homogeneous generators of ${\cal R}$ may not have the same degree and the ambient space is then a weighted projective space. Further, given a ring ${\cal R}$ with an ideal ${\cal I}$, we have the following short exact sequence
\begin{equation}
0\rightarrow{\cal I}\rightarrow{\cal R}\rightarrow{\cal R}/{\cal I}\rightarrow 0\ .
\end{equation}
For a reductive group action $G$ on ${\cal R}$ we can treat the ring as a $G$-module therefore we have \cite{MelvinHochster}:
\begin{equation}
0\rightarrow {\cal I}^G\rightarrow{\cal R}^G\rightarrow ({\cal R}/{\cal I})^G\rightarrow 0\ .
\end{equation}
We require ${\cal I}$ to be invariant under $G$, and therefore we have
\begin{equation}
({\cal R}/{\cal I})^G = {\cal R}^G/{\cal I}\ .
\end{equation}
We are interested in the case where $V\subset\mathbb{P}^N$ and ${\cal I} = I(\widetilde{V})$, ${\cal R} = R(\mathbb{P}^N) = A(\mathbb{C}^{N+1})$.
We therefore have the following exact sequence
\begin{align}
0\rightarrow I(\widetilde{V})^G\rightarrow \mathbb{C}[x_0,x_1,x_2,\cdots,x_N]^G\rightarrow \left(\frac{\mathbb{C}[x_0,x_1,x_2,\cdots,x_N]}{I(\widetilde{V})}\right)^G\rightarrow 0\ ,
\end{align}
where $I(\widetilde{V})^G = I(\widetilde{V})$ which leads to
\begin{equation}
A(\widetilde{V})^G = \frac{\mathbb{C}[x_0,x_1,x_2,\cdots,x_N]^G}{I(\widetilde{V})} = \frac{\mathbb{C}[y_0,y_1,y_2,\cdots,y_M]/S}{I(\widetilde{V})} = \frac{\mathbb{C}[y_0,y_1,y_2,\cdots,y_M]}{\langle S,I(\widetilde{V})\rangle}\ ,
\end{equation}
where $\langle y_0,y_1,y_2,\cdots,y_M\rangle$ is the set of $G$-invariant polynomials in $\mathbb{C}[x_0,x_1,x_2,\cdots,x_N]$ and $S$ is the ideal of the relations between them, {\it {\it i.e.}}, the kernel of the map (\ref{eq:relations_invpolys}). As the final step we will take $\text{Proj}(A(\widetilde{V})^G)$ which defines the quotient of $V$ as a projective subvariety in a projective ambient space. We emphasize again that the ambient space could be a weighted projective space. In the above calculation we write $I$ in terms of $\mathbb{C}[y_0,y_1,y_2,\cdots,y_M]$, {\it {\it i.e.}}, we do the following coordinate transformation
\begin{equation}
f(x_0,x_1,x_2,\cdots,x_N) = F(y_0,y_1,y_2,\cdots,y_M)\ ,
\end{equation}
for $I = \langle f\rangle$. To be more precise, in the intermediate step we should really write $I(\widetilde{V})$ in terms of the generators of the ring $\mathbb{C}[y_0,y_1,y_2,\cdots,y_M]/S$ and the ideal $I$ in the last step is actually the inverse image of the intermediate $I(\widetilde{V})$ under the map
\begin{align}
\pi: \mathbb{C}[y_0,y_1,y_2,\cdots,y_M]\rightarrow \mathbb{C}[y_0,y_1,y_2,\cdots,y_M]/S\ .
\end{align}

For our purpose, $V = \overline{C}$, $f$ is the defining equation in $\mathbb{P}^2$ whose homogeneous coordinate ring is $R = \mathbb{C}^h[X,Y,Z]$ and the $G$-action is the uplift of the $r$-action on $\mathbb{C}^*\times \mathbb{C}^*$ to $\mathbb{P}^2$. In particular we see that the $GL(3,\mathbb{C})$ matrix $M_r$ naturally provides a lift of the $G$-action on $\mathbb{P}^2$ to $\mathbb{C}^3$. Therefore effectively we will be studying a $G$-action on $\mathbb{C}^3$. Viewed in these terms, our surface $S$ corresponds to the variety $\mathbb{C}^3/G$. For our $\mathbb{Z}_3$-action, $S$ can be embedded into a weighted projective space with coordinate ring $\mathbb{C}[w,x,y,z]$. $S$ corresponds to the coordinate ring $\mathbb{C}[w,x,y,z]/I(\widetilde{S})$ after going over to the affine cone. Here $\langle w,x,y,z\rangle$ are invariant polynomials of the $\mathbb{Z}_3$-action. We can also write $I(\widetilde{C})$ in terms of $\langle w,x,y,z\rangle$. Therefore in this case we have:
\begin{equation}
A(\widetilde C)^G = \frac{\mathbb{C}[w,x,y,z]}{\langle I(\widetilde{S}),I(\widetilde{C})\rangle}.
\end{equation}
Hence we have:
\begin{equation}
C^G:=\text{Proj}\left(\frac{\mathbb{C}[w,x,y,z]}{\langle I(\widetilde{S}),I(\widetilde{C})\rangle}\right).
\end{equation}
In our cases $\text{deg}(w,x,y,z) = (1,2,3,3)$, and we have $C^G = S\cap C\subset\mathbb{P}^{1,2,3,3}$. We will give the exact forms of $\langle w,x,y,z \rangle$ in terms of $\langle X,Y,Z \rangle$, and $I(\widetilde{S})$ and $I(\widetilde{C})$ in terms of $\langle w,x,y,z \rangle$ in a moment when we present a concrete example.

Physically the above derivation shows that starting from $(\mathbb{C}^*\times\mathbb{C}^*, C)$ after a $\mathbb{Z}_3$ action, we have arrived at the configuration $(S, C^G)$. In particular, we have shown that one can conveniently describe the pair $(S, C^G)$ by an embedding into a (weighted projective space) $C^G = S\cap C \subset S \subset w\mathbb{P}$. This description in terms of embeddings into a weighted projective space is useful for the computation of the physical data of the corresponding gauge theories, as we will now see. In particular, the gauge symmetry is determined solely by $C^G$, and as a zeroth-order consistency check we compute its genus.

For a concrete example we compute the $N=3$ case. The invariant polynomials of $G$ on $\mathbb{C}^3$ are:
\begin{equation}\label{eq:invpolys_G}
\begin{split}
w &= X + Y + Z, \\
x &= X^2 + Y^2 + Z^2, \\
y &= X^3 + Y^3 + Z^3, \\
z &= XY^2 + YZ^2 + ZX^2 ,
\end{split}
\end{equation}
subject to the relation
\begin{equation}\label{eq:S_eq}
27 w^2 x^2-9 w^4 x+8 w^3 y+w^6-48 w x y-24 w x z-3 x^3+24 y^2+24 y z+24 z^2 = 0 ,
\end{equation}
for the surface $S\subset\mathbb{P}^{1,2,3,3}$. As we have anticipated we have $\text{deg}(w,x,y,z) = (1,2,3,3)$.

We then consider the compactification to $\mathbb{P}^2$ of $C_{T_3}\subset\mathbb{C}^*\times\mathbb{C}^*$ defined by:
\begin{align}
I(\widetilde{C})=\,\,&c_1 \left(X^3+Y^3+Z^3\right)+ c_2 \left(X^2 Z+X Y^2+Y Z^2\right)\notag\\
&+ c_3\left(X^2 Y+X Z^2+Y^2 Z\right)+\ c_4 X Y Z = 0 .
\end{align}
Writing $I(\widetilde{C})$ in terms of the invariant polynomials \ref{eq:invpolys_G}, we have
\begin{equation}\label{eq:C_eq}
c_1 y + c_2 z + c_3 (w x-y-z) + \frac{1}{6} c_4\left(w^3-3 w x+2 y\right) = 0 .
\end{equation}
We then take the $w = 1$ patch of $\mathbb{P}^{1,2,3,3}$ which is isomorphic to $\mathbb{C}^3$. On this patch we solve Eq. \ref{eq:S_eq} for $z$ and plug it into Eq. \ref{eq:C_eq} and we have the following curve $\mathcal{C}\subset\mathbb{C}^2$:
\begin{align}
\begin{split}
0 =\ & 6 c_2 (2 (x-y) (6 c_1 Y+c_4 (-3 x+2y+1)) + c_3 (-3 x^3+27 x^2-48 xy -9x +24 y^2+8 y+1)) \\
& + 12c_3 (x-y) (6 c_1 y+c_4 (-3 x+2 y+1))+2 (6 c_1y+c_4 (-3 x+2 y+1))^2 \\
& + 3 c_2^2 (-3 x^3+27 x^2-48 x y-9 x+24y^2+8 y+1) \\
& +3 c_3^2 (-3x^3+27 x^2-48 x y-9 x+24 y^2+8y+1) .
\end{split}
\end{align}
Now it is easy to check that $g(\mathcal{C}) = 1$. We could perform the same calculation for the other cases and we obtain the series $g(\mathcal{C}) = 1, 1, 2, 4, 5, 7, 10, 12, 15, 19, \dots$ for $N = 3, 4, 5, 6, 7, 8, 9, 10, 11, 12,...\,$. These results agree with the ranks for the corresponding gauge groups computed above by several other methods, giving us confidence in the above procedure.

\section{Conclusion and future directions}\label{sec:conclusions}

In this paper, we have discussed the five-dimensional superconformal field theories that arise from M-theory on a ${\mathbb C}^3/\Delta(3N^2)$ singularity. We conjectured the existence of new theories and studied them from the point of view of crepant resolutions within M-theory and also through S-folds of 5-brane webs.


Starting from the 5-brane web description of the $T_N$ theory, the additional ${\mathbb Z}_3$ orbifold action contained in $\Delta(3N^2)$ is interpreted as an S-fold {\it via} an element of $SL(2,{\mathbb Z})$ which maps the mutually non-local branes into each other. Geometrically this acts as a $2\pi/ 3$ rotation in the 56-plane transverse the branes. This in turn introduces a conical deficit at the origin and hence additional 7-branes that carry an $E_6$ flavour symmetry. It also breaks the original $SU(N)^3$ flavour symmetry of the $T_N$ theory to $SU(N)$. The resulting theories split into two classes depending on whether or not the 5-branes intersect the 7-brane at the origin. For $N\ne 3k$, they do, and the $E_6$ is broken to $SU(3)$, whereas for $N=3k$ they do not and $E_6$ is preserved. This leads to five-dimensional conformal field theories with $E_6\times SU(N)$ or $SU(3)\times SU(N)$ flavour symmetries (which, for $N=3,4$, are further enhanced to $E_8$ and $E_5$ respectively). Note that the effect of gauging the ${\mathbb Z}_3$ symmetry is not to simply induce a renormalization group flow, as one might expect in more traditional gaugings. Rather the inclusion of  7-brane at the origin can give rise to additional degrees of freedom not present in the parent theory.  

Besides the evidence we have shown in this paper, it will be interesting to give an explicit complete crepant resolution of $X=\mathbb{C}^3/\Delta(3N^2)$ and calculate the intersection numbers to study the structure of the ECB of $\mathcal{T}_5\{X\}$. It would also be very interesting to study the Higgs branch of $\mathcal{T}_5\{X\}$ where one has to study the (normalizable) deformations of the non-isolated singular locus in $X$. Also, as $\Delta(3N^2)$ is not the only finite subgroup of the $SU(3)$ quotient by which leads to a canonical singularity that admits a crepant resolution. Hence potentially corresponds to a five-dimensional SCFT upon M-theory compactification, one may also study the other orbifolds $X=\mathbb{C}^3/G$ where $G$ is a finite subgroup of $SU(3)$ and the corresponding five-dimensional theory $\mathcal{T}_5\{X\}$.

%
%

\section*{Acknowledgements}

We thank Francesco Benini, Sergio Benvenuti, Yukari Ito, Sakura Schafer-Nameki, Benjamin Sung and Yi-Nan Wang for helpful discussions. The work of BA, MN, and JT is supported by a grant from the Simons Foundation (\#488569, Bobby Acharya).

\bibliographystyle{JHEP}
\bibliography{F-ref}

\end{document}